\documentclass[reprint,
superscriptaddress,
showpacs,
preprintnumbers,
nofootinbib,
nobibnotes,
amsmath,
amssymb, 
aps,
prd,
onecolumn,
notitlepage,
floatfix]{revtex4-1}
\usepackage[dvipdfmx]{graphicx}

\usepackage[utf8]{inputenc}
\usepackage[normalem]{ulem}

\usepackage{multirow,relsize,slashed,textpos}
\usepackage{wasysym}
\usepackage{marvosym}
\usepackage{tikz,soul}
\usepackage{braket} 
\usepackage{mathrsfs}
\usepackage{xspace}
\usepackage{bm}
\usepackage[pdfencoding=auto,psdextra,colorlinks=true,allcolors=purple]{hyperref}

\newcommand{\refeq}[1]{Eq.~(\ref{#1})}
\newcommand{\refeqs}[2]{Eqs.~(\ref{#1})~and~(\ref{#2})}

\newcommand{\reffig}[1]{Fig.~\ref{#1}}
\newcommand{\reffigs}[2]{Figs.~\ref{#1}~and~\ref{#2}}
\newcommand{\refsec}[1]{Section~\ref{#1}}
\newcommand{\refapp}[1]{Appendix~\ref{#1}}
\newcommand{\reftab}[1]{Table~\ref{#1}}
\newcommand{\refref}[1]{Ref.~\cite{#1}}

\renewcommand{\phi}{\varphi}

\newcounter{CommentCount}
\definecolor{MH}{rgb}{0.0,0.6,9}

\begin{document} 
\preprint{\hfill FTPI-MINN-20-07}

\title{Pair production of dark particles in meson decays}

\author{Matheus Hostert}
\email{mhostert@umn.edu}
\affiliation{School of Physics and Astronomy, University of Minnesota, Minneapolis, MN 55455, USA}
\affiliation{William I. Fine Theoretical Physics Institute, School of Physics and Astronomy, University of
Minnesota, Minneapolis, MN 55455, USA}

\affiliation{Perimeter Institute for Theoretical Physics, Waterloo, ON N2J 2W9, Canada}

\author{Kunio Kaneta}
\email{kkaneta@umn.edu}
\affiliation{School of Physics and Astronomy, University of Minnesota, Minneapolis, MN 55455, USA}
\affiliation{William I. Fine Theoretical Physics Institute, School of Physics and Astronomy, University of
Minnesota, Minneapolis, MN 55455, USA}

\author{Maxim Pospelov}
\affiliation{School of Physics and Astronomy, University of Minnesota, Minneapolis, MN 55455, USA}
\affiliation{William I. Fine Theoretical Physics Institute, School of Physics and Astronomy, University of
Minnesota, Minneapolis, MN 55455, USA}

\preprint{}
\begin{abstract}
Rare decays of $K$ and $B$ mesons provide a powerful probe of dark sectors with light new particles. We show that the pair production of $O(100\,{\rm MeV})$ dark states can be probed with the decays of $K_L$ mesons, owing to the enhanced two-body kinematics, $K_L\to X_1X_2$ or $X_2X_2$. If either of these two particles is unstable, {\em e.g.} $X_2\to X_1\pi^0$, $X_2\to X_1\gamma$ or $X_{1,2}\to \gamma\gamma$, such decays could easily mimic $K_L\to \pi^0 \nu\overline{\nu}$ signatures, while not being ruled out by the decays of charged kaons. We construct explicit models that have enhanced $K_L$ decay signatures, and are constrained by the results of the KOTO experiment. We note that recently reported excess events can also be accommodated while satisfying all other constraints ($B$ decays, colliders, beam dumps). These models are based on the extensions of the gauge and/or scalar sector of the theory. The lightest of $X_{1,2}$ particles, if stable, could constitute the entirety of dark matter. 
\end{abstract}
\maketitle

\section{Introduction}

Long lived mesons such as neutral and charged $K$ and $B$ have long been used as sensitive probes of new physics. New sources of flavor change can be constrained
at scales much exceeding the direct reach of present and future colliders \cite{Wolfenstein:1964ks}.
At the same time, flavor physics constrains many well-motivated models with light new states, such as {\em e.g.} axions, axion-like particles, dark photons and scalars coupled through Higgs portal, see {\em e.g.} \cite{Beacham:2019nyx}.

Given significant experimental advances in flavor physics, and its planned expansion, we would like to revisit its sensitivity to the pair production of 
light new particles. A prominent example of that nature was introduced in \cite{Bird:2004ts,Bird:2006jd}, 
where the pair production of sub-GeV dark matter particles in $B$ meson decays was studied (see also \cite{Badin:2010uh}). Relevant signatures included $B\to K^{(*)}+2X$, where $X$ is some dark state.  Generalizing such channels to a pair of non-identical states, we would like to consider
\begin{align}
    K,B \to X_{1} \,+\, X_2 + Y_{\rm SM}
\end{align}
signatures where $X_{1,2}$ stand for new exotic particles, while $Y_{SM}$ is a
variety of possible Standard Model (SM) states accompanying the decay. 

The pair production of dark states offers a certain edge to the decays of \emph{neutral} $K$ and $B$ mesons. For the decay of neutral mesons, $Y_{\rm SM}$ can be exactly \O, while for the charged mesons $Y_{\rm SM}$ would have to carry electric charge, and therefore consists of at least one SM particle. Thus, for example, an underlying $s-d-X_1-X_2$ amplitude is expected to induce $K_L\to X_1X_2$ decays that are faster than 
$K^+\to \pi^+ X_1X_2$, and the latter could be even energetically forbidden. Moreover, if the generalized current producing $X_{1,2}$ is ``nearly conserved", 
that is its non-conservation is controlled by relatively small mass parameters $\propto m_X$,
then $B^0\to X_1X_2$ can also be generically suppressed. 

Of course, if both $X_{1,2}$ are stable, then the $K^0\to X_1X_2$ process would not
result in strong bounds on dark sectors, as fully invisible decays of neutral kaons 
are difficult to probe experimentally (\emph{cf.} Refs.~\cite{McElrath:2005bp,McKeen:2009rm,Kamenik:2011vy,Dreiner:2009er,PhysRevD.91.015004,
PhysRevD.92.034009,PhysRevD.98.035049}). This situation changes if one or both of the 
$X$ particles is unstable, producing SM particles in the decay. In particular, the production of $\pi^0$ is of special interest, as it fits the SM signature of $K_L \to \pi^0 \nu\overline{\nu}$ decay, that is being searched for in an existing experiment 
at J-PARC, KOTO, \cite{Ahn:2018mvc}, and has been proposed as a motivation for the CERN-based experiment, KLEVER \cite{Moulson:2019ifj}. To be concrete, we will analyse the following signature:
\begin{align}
\label{pi0production}
    K_L \to X_{1} \,+\, X_2& \\\nonumber
    &^\searrow\, \pi^0 \,+\, X_1,
\end{align}
within some broad class of models of dark sectors. Given limited reconstruction capabilities for four-momenta of photons, the $\pi^0\nu\overline{\nu}$ signature can also be mimicked by an exotic particle decay to photons. 
In particular, we find that in some models the following signature is promising: 
\begin{align}
\label{2gproduction}
    K_L \to X_{1} \,+\, X_2& \\\nonumber
    &\;\;^\searrow\, \gamma\gamma \,+\, X_1,
\end{align}
Finally, both particles can be unstable, giving a single photon in the decay,
\begin{align}
\label{dipoleproduction}
    K_L \to X_{2}& \,+\, X_2 \\\nonumber
    ^\searrow\,&  X_1+\gamma ^\searrow\, X_1+\gamma.
\end{align}
The main question for us to study is the following: could some minimal models of dark sector lead to the above signature, so that one should expect the decays of $K_L$ - as opposed to the decays of $K^\pm$ and $B$ mesons - be the leading probe of such models?

\begin{figure*}
    \centering
    \begin{tabular}{ccc}
    \includegraphics[width=0.249\textwidth]{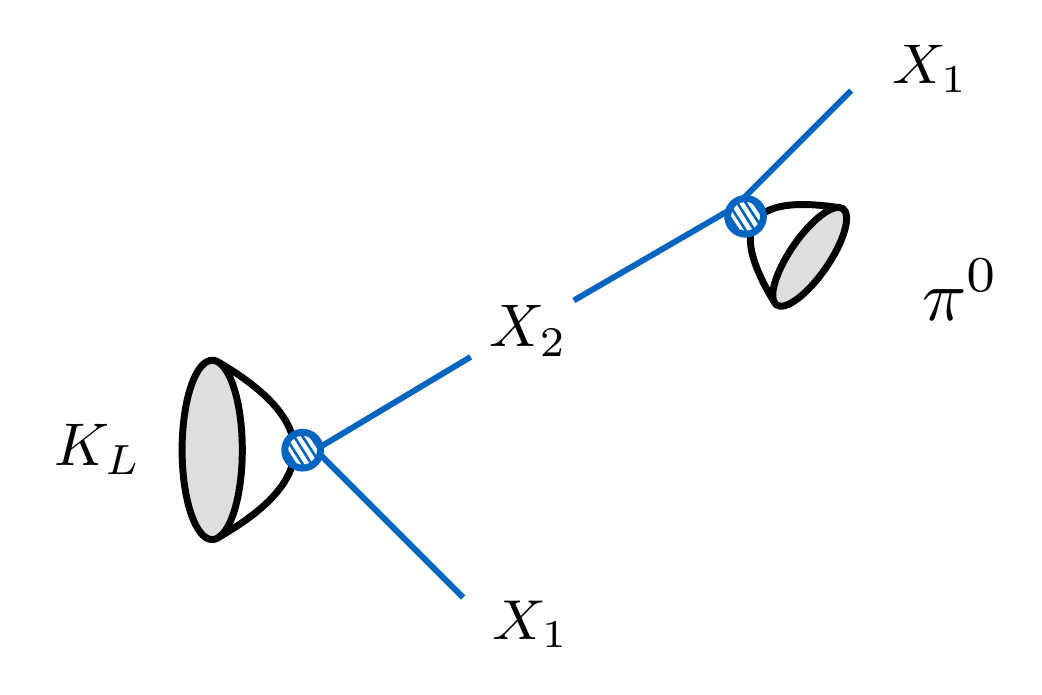}
    \includegraphics[width=0.249\textwidth]{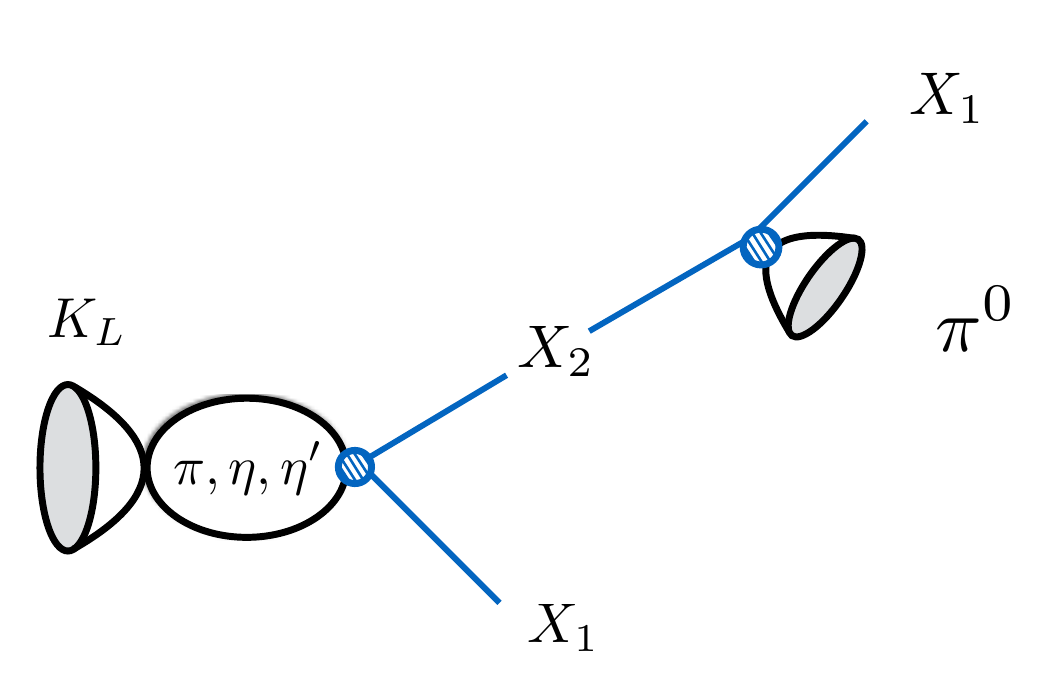}
    &   
    \includegraphics[width=0.249\textwidth]{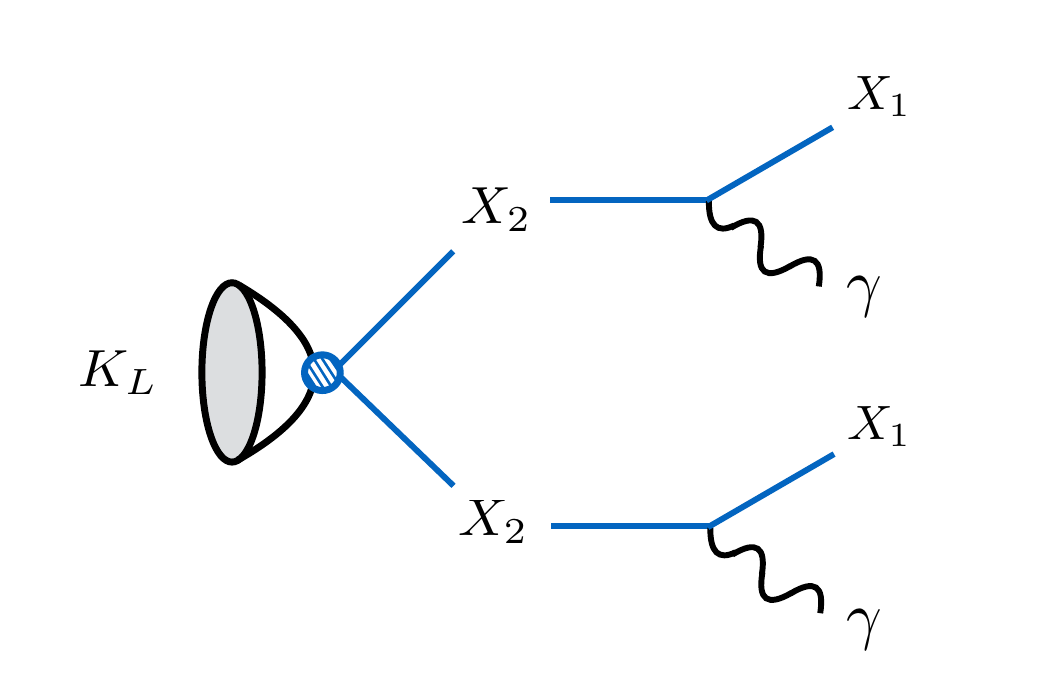}
    &
    \includegraphics[width=0.249\textwidth]{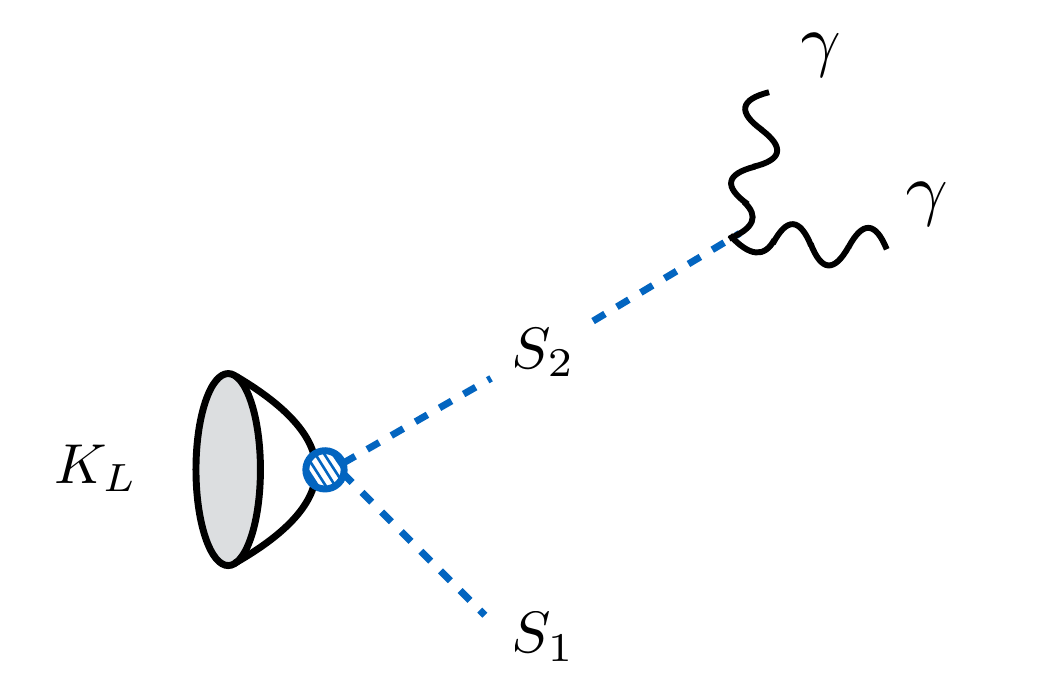}
    \\
    A) \textsc{$\pi^0$ production}
    & 
    B) \textsc{dipole portal}
    &
    C) \textsc{$\pi^0$ impostor}
    \end{tabular}
    \caption{The $K_L$ decay to an arbitrary neutral sector ($X_1$ and $X_2$), followed by their subsequent decay in three different scenarios. Pair production takes place via a heavy mediator shrunk to a point in the diagrams. For $\pi^0$ production, we propose two scenario: FCNC via new mediators and a long-distance $\Delta S = 1$ transition followed by a flavor diagonal coupling to a new mediator.}
    \label{fig:diagrams}
\end{figure*}

In this paper we show that the answer to this question is affirmative, and present 
several scenarios, based on vector and scalar portal models, that lead to 
measurable rates of $K_L$ decays, exceeding the SM rate for $K_L \to \pi^0 \nu\overline{\nu}$, while evading the bounds from collider, beam dump, and flavor probes. The topologies we consider are shown in \reffig{fig:diagrams}, where typically a new heavy portal particle can be integrated out and mediates $K_L\to X_1 X_2$ production from either SM-like flavor changing neutral current (FCNC) couplings or via flavor diagonal couplings attached to long distance $\Delta S = 1$ operators.
The new FCNC couplings are built using minimal flavor violation (MFV) ansatz, where all sources of flavor change originates from the SM Yukawa matrices 
(see {\em e.g.} \cite{DAmbrosio:2002vsn}). 
The MFV framework also allows to connect, in most models, $X_1-X_2-\pi^0$ vertex 
with the corresponding vertex of $\eta$ meson, resulting in the new decay channels,
$\eta\to X_1X_2$. Subsequent decay of $X_2$ away from the production point puts 
strong constraints on these models from the results of the beam dump searches at highest energies. 
We also show that for models responsible for processes (\ref{pi0production}) and (\ref{dipoleproduction}) the lightest of two dark states, $X_{1}$, can be stable, and therefore contribute to dark matter. However, this typically requires additional components to such models, that are not probed directly by $K$ and $B$ physics. 

Our paper is inspired, in part, by a recent report by the KOTO collaboration that faces four unexplained events after unblinding their data, at the level much larger than the SM neutrino channel, which prompts theoretical investigations of beyond-SM (BSM) 
physics that may lead to such a signature. Many recent studies have appeared, where the focus has been primarily on $K_L$ decays to pion plus new invisible particles~\cite{Kitahara:2019lws,Fabbrichesi:2019bmo,Egana-Ugrinovic:2019wzj,Dev:2019hho,Li:2019fhz,Liu:2020qgx,Jho:2020jsa,Cline:2020mdt,Ziegler:2020ize}. Other recent alternatives include new particles produced at the target that decay to $\gamma\gamma$ inside the KOTO detector~\cite{Kitahara:2019lws}, direct $K_L$ decay to $\gamma\gamma$ plus invisible states~\cite{Liao:2020boe}, and heavy new physics operators with flavor violation or $\Delta I = 3/2$ structure~\cite{Kitahara:2019lws,Pich:2020gan,He:2020jzn,He:2020jly}. 
In this work we take a different approach and address whether the processes (\ref{pi0production}), (\ref{2gproduction}), and (\ref{dipoleproduction}) could be behind KOTO events, passing all the experimental requirements including the distribution over transverse momentum $p_T$. This differs from previous studies due to the complete annihilation of $K_L$ to a dark sector, creating new avenues to fake $K_L \to \pi^0\slashed{E}$ signatures.

Another strong motivation for us is the upcoming ultra-high luminosity Belle II experiment \cite{Kou:2018nap}, where significant progress with measurements of $B$ meson decays accompanied by missing energy is expected. Given that our models are built using the MFV framework, direct connection between $K$ and $B$ meson decays can indeed be established. 

This paper is organized as follows: in the next section we give a brief overview of experimental situation regarding the neutrino pair production decays of the charged and neutral $K$ and $B$ mesons. In Section III, we construct $s-d-X_1-X_2$ and $b-s-X_1-X_2$ amplitudes that result from a simple vector and Higgs portal models. 
In section IV, we calculate observable consequences for meson decays, including the $p_T$ distributions of photons in KOTO setting, and constrain parameters of these models. We also analyze the suitability of these models as an explanation of KOTO events, including the overall rate and the distribution over $p_T$. In Section V, we construct explicit models, where the lightest of the two particles, $X_1$, is in fact the dark matter, passing all existing constraints. We reach our conclusion in Section VI.  

\section{Experimental prospects for \(\nu\overline{\nu}\) modes in \(K,B\) decays}

The decays of $K_L$ and $K^\pm$ mesons to a pion and $\nu\overline{\nu}$ pair are of special importance for the precision tests of the SM. The absence of long-distance contributions to the amplitudes, and the simplicity of the resulting $s-d$ vector current matrix element offers a perfect testing ground for the CKM paradigm (see, {\em e.g.} \cite{Buchalla:2008jp}). While $K^+\to \pi^+\nu\overline{\nu}$ decay has been
observed with a handful of events \cite{Artamonov:2008qb}, only the upper limits exists for the neutral mode, $K_L\to \pi^0\nu\overline{\nu}$ \cite{Ahn:2018mvc}. 

As is well known, the FCNC are forbidden at tree level in the SM, but are generated at one loop due to $Z$-penguin and $W$-box diagrams. In fact, the underlying $s \leftrightarrow d$ transitions for $K \to \pi \overline{\nu}\nu$ processes have been the object of several studies~\cite{Buchalla:1993wq,Buras:1998raa}, where the SM prediction can be easily calculated using the effective Hamiltonian. Neglecting the up quark contribution, 
\begin{equation}
\label{SM}
    \mathcal{H}_{\rm eff} = \frac{\sqrt{2} G_F \alpha}{\pi\sin^2{\theta_W}} \left(V_{ts}^*V_{td} X(x_t) + V_{cs}^*V_{cd} X(x_c)\right) \left( \overline{s} \gamma_\mu P_L d\right) \left( \overline{\nu} \gamma^\mu P_L \nu \right) \,+\,{\rm h.c.},
\end{equation}
where $x_i=m_i^2/M_W^2$ and $X(x_i)$ stands for a well-known loop function. The fairly robust SM theoretical predictions currently are~\cite{Buras:2015qea}
\begin{align}
\mathrm{BR} \left( K^+ \to \pi^+ \overline{\nu} \nu \right) &= \left(0.84 \pm 0.10 \right) \times10^{-10},
  \\  \mathrm{BR} \left( K_L \to \pi^0 \overline{\nu} \nu \right) &= \left(0.34 \pm 0.06 \right) \times10^{-10}.
\end{align}
The measurement of the charged mode will soon be refined by the on-going NA62 
experiment \cite{CortinaGil:2018fkc}, which may reach $\sim$10\% accuracy in measuring this branching ratio. The neutral mode is currently being pursued by the 
KOTO collaboration \cite{Ahn:2018mvc}, while new more sensitive experiments are being planned \cite{Moulson:2019ifj}. 

These kaon decay modes, so much suppressed in the SM, can serve as a powerful probe of physics beyond SM. Indeed, short-distance new physics can alter the Wilson coefficients in (\ref{SM}) and enhance (or in case of the destructive interference, suppress) the corresponding decay rates. Nevertheless, the considerations of the isospin invariance lead to the so-called Grossman-Nir \cite{Grossman:1997sk} bound that restricts the neutral mode 
relative to the charged mode, 
\begin{equation}
\label{GN}
   \mathrm{BR} \left( K_L \to \pi^0 \overline{\nu} \nu \right) < 4.4 \times  \mathrm{BR} \left( K^+ \to \pi^+ \overline{\nu} \nu \right).
\end{equation}

Taking into account this bound, and the existing measurement of the charged mode, one should not expect any signal in the neutral mode
at the current level of sensitivity of KOTO, regardless whether short-distance new physics exists. Given this consideration, the recent KOTO report of anomalously high number of $K_L \to \pi^0 \nu\overline{\nu}$ events~\cite{kotoKAON19} stands out. A total of 4 events are observed, being only one of them consistent with background expectations. The 3 anomalous events imply a total BR of~\cite{Kitahara:2019lws}
\begin{equation}
\mathrm{BR} \left( K_L \to \pi^0 \overline{\nu}\nu \right) = \left(2.1^{+4.1}_{-1.7} \right) \times10^{-9} \,\, \text{at KOTO.}
\end{equation}

It is clear that if the nature of these anomalous events is clarified, and understood to be {\em not} associated with unaccounted backgrounds, it would require some special type of new physics, unlikely to be associated with the short distance modification of $d-d-\nu-\nu$ amplitude. In the past, it was emphasized that should 
light sub-$m_K$ new physics states exist, they can significantly alter the relation between neutral and charged mode \cite{Fuyuto:2014cya,Hou:2016den}.

Analogous $B$ mesons decays involving missing energy are more challenging, both due to the hadronic form factors input in the theory prediction, but also due low efficiency of such searches that rest upon a full reconstruction of 
one of the $B$ mesons at $B$ factories.
For $B\to K\overline{\nu}\nu$, the current predictions stand at~\cite{Buras:2014fpa}
\begin{equation}
   \mathrm{BR} \left( B^+ \to K^+\overline{\nu}\nu \right) = (4.0\pm0.5)\times10^{-6}.
\end{equation}
The current best limits on these decays come from Belle~\cite{Grygier:2017tzo}, and are given at 90\% C.L. as 
\begin{equation}
   \mathrm{BR} \left( B \to K^{\phantom{+}}\!\!\overline{\nu}\nu \right)  < 1.6\times10^{-5} \,\,\text{at Belle},
\end{equation}
from a combination of neutral and charged $B$ decays. Whether such limits constrain the production of new light particles in kaon decays is a highly model dependent question, but for models with SM-like FCNC, $B$ decays present somewhat weaker but still competitive bounds. Indeed, given the upcoming programme at Belle-II, it is timely to consider these channels. At full luminosity, Belle-II will be able to measure the missing energy $B$ decays to within $10\%$ precision~\cite{Kou:2018nap}, which together with new dedicated searches, will bring about significant improvement on dark sectors limits below $\sim 5$ GeV.

\section{Survey of Models and FCNC Amplitudes}\label{sec:models}

Meson decays to a pair of light dark states at the quark level implies a 
higher-dimensional effective operators, when the ``mediator" physics is heavy and integrated out. In this section we give example of such operators, and show the pathways of their emergence via exchange of the exotic scalar and vector particles from UV complete theories.  Our goal for this section is to determine a selection of representative models, preferably with UV completion, that could give enhanced $K_L$ decay signatures to dark states. We summarize the set of such models at
the end of this section, in subsection E and Table \ref{tab:model_summary}.

\subsection{Effective operators in the MFV approach}

The main operator structures that we shall consider will be 
\begin{equation}
\label{operators}
    O^V_{sd}  = g^V_{sd} (\overline{s_L} \gamma_\mu d_L) \times J_X^\mu;\qquad
    O^S_{sd}  = g^S_{sd} m_s(\overline{s_R} d_L) \times J_X,
\end{equation}
where $g^{V(S)}_{sd}$ are the coefficients that package together all 
the physics responsible for $s\to d$ transitions, as well as details of mediators of interaction with $X$-sector. $J_X^\mu$ and $J_X$ are some generalized currents that transform as vectors and scalars, and are built from the $X_1$ and $X_2$ fields. At this point, we do not assume anything about the properties of $X_{1(2)}$ other than that they can be pair produced on shell in $K_L$ decays. Possible forms for such ``dark" currents can be easily listed: 
\begin{align}
\label{Xboson}
    {\rm bosons:}&&\,\, J_X^\mu &= X_1\partial^\mu X_2 - (\partial^\mu X_1) X_2,...,&&J_X = X_1^2,\, X_2^2,\, X_1X_2,..., \\
    {\rm fermions:}&&\,\, J_X^\mu &= \overline{X_1}\Gamma^\mu X_2 + \overline{X_2}\Gamma^\mu X_1,...,&&J_X = \overline{X_1}X_1,\, \overline{X_2}X_2,\, \overline{X_1}X_2 + \overline{X_2} X_1,...,
\label{Xfermion}
\end{align}
where $\Gamma^\mu = \gamma^\mu,\,\gamma^\mu\gamma_5,\, i\sigma_{\mu\nu}\partial_\nu$ etc. Scalar fermionic currents can also include pseudoscalar combinations, $\overline{X_1} i\gamma_5 X_1$. 

Note that although these operators look fairly general, they do not represent
an exhaustive set. Indeed, at the effective level, one could imagine 
{\em i.e.} the presence of $s-d$ right-handed currents, and/or scalar 
currents that emerge without chiral suppression $\propto m_s$. Our choice is motivated by the SM-like mechanisms for the FCNCs, that specifically operate 
with left-handed light quark fields, and require mass (or Yukawa) insertion whenever $q_L$ chirality is flipped to $q_R$. 

Another feature of the SM-like FCNC's is the adherence to the MFV ansatz, which is a powerful framework that allows to connect $g^{V(S)}_{sd}$ couplings with those that involve $b\to s$ transitions, as only the Yukawa matrices of the SM source flavor transitions. We assume that $X$-currents do not transform under SM flavor rotations, and take the following 
MFV structures that give rise to our effective flavor-changing operators, 
\begin{equation}
 g^V_{sd} (\overline{s_L} \gamma_\mu d_L)   \subset  a \overline{Q_L} Y_U Y_U^\dagger \gamma_\mu Q_L;\qquad g^S_{sd} m_s(\overline{s_R} d_L) \subset b \overline{D_R} M_D^\dagger Y_U Y_U^\dagger Q_L.
 \label{MFV}
\end{equation}
These structures are parametrized now by just two (complex) coefficients 
$a$ and $b$, which leads to rigid relations between $s\to d$ and $b \to d$ transitions induced by these structures. Moreover, given the dominance of the top-quark and charm-quark Yukawa couplings, we can effectively reduce flavor-changing coefficients to the corresponding products of the CKM matrix elements, 
\begin{equation}
  g^V_{sd} = a  (y_t^2 V_{ts}^*V_{td}+y_c^2 V_{cs}^*V_{cd}); \qquad  g^S_{sd} = b (y_t^2 V_{ts}^*V_{td}+y_c^2 V_{cs}^*V_{cd}),
\end{equation}
where $y_{t,c}$ stand for the Yukawa coupling of the top and charm quark.

One can make further progress in general, without explicitly defining 
UV completions, and exploiting the $CP$-properties of the operators. 
To that end, let us assume that $X$-currents are self-conjugate, 
$J_X^\dagger =J_X$, as is the case for all examples of Eqs. (\ref{Xboson}) and (\ref{Xfermion}). This condition will be automatically satisfied if operators (\ref{operators}) are induced by an exchange of 
the real scalar or vector field, $S$ and $Z'$. Moreover, if couplings 
of $S$ and $Z'$ to SM particles are $CP$-even, then $a$ and $b$ are real.
These properties will allow us to establish whether the corresponding $K_L$ 
amplitudes are given by real or imaginary part of the product of 
CKM matrix elements. Neglecting small, $\epsilon_K$-sized admixture, 
$K_L$ coincides with the $CP$-odd combination of neutral kaons, 
$K_L = 2^{-1/2}(K^0-\overline{K^0})$, where $CP(K^0) = \overline{K^0}$ and vice versa. 
With this definition, and Eqs. (\ref{operators}) and (\ref{MFV}), 
we can determine which combination of the CKM matrix elements is responsible for a given transition,
\begin{align}
    \langle 0 | O^V_{sd}| K_L &\rangle
    \propto a{\rm Re} (y_t^2 V_{ts}^*V_{td}+y_c^2 V_{cs}^*V_{cd}), &&\langle 0 | O^S_{sd}| K_L \rangle
    \propto b{\rm Im} (y_t^2 V_{ts}^*V_{td}+y_c^2 V_{cs}^*V_{cd}), \\
    \langle \pi^0 | O^V_{sd}| K_L &\rangle
    \propto a{\rm Im} (y_t^2 V_{ts}^*V_{td}+y_c^2 V_{cs}^*V_{cd}), &&\langle \pi^0 | O^S_{sd}| K_L \rangle
    \propto b{\rm Re} (y_t^2 V_{ts}^*V_{td}+y_c^2 V_{cs}^*V_{cd}).
\end{align}
Note that top quark provides a dominant contribution, especially in Im$(...)$.
These relations arise from specific $C$, $P$ and $CP$ properties of operators that mediate the transition. For the processes considered in this paper, 
$K_L\to X-{\rm pair}$, the quark bilinears that mediate the transition 
between $K_L$ and the vacuum are $C$-even and $CP$-odd combinations $\overline{d}\gamma_\mu\gamma_5s + \overline{s}\gamma_\mu\gamma_5 d$ for the vector type operator, and $\overline{d}i\gamma_5s + \overline{s}i\gamma_5 d$ for the scalar type. 

\subsection{UV completion via Higgs portal}

In this subsection, we give some examples of UV completions for operators
(\ref{operators}) using Higgs portals. From now on, we will stick to the notation where $X_i=S_i$ if it is a scalar particle, and $X_i=\psi_i$ if a fermion. Choosing the minimal Higgs content, we can couple $H^\dagger H$ Higgs field bilinear to terms that are linear or quadratic in terms of the ``dark" scalar fields $S_i$. Writing this fields in the real scalar field basis, we get
\begin{equation}
\label{Hportal1}
    {\cal L}_{\rm H-portal} = H^\dagger H \times \left( \sum_{i} A_i S_i + \sum_{ij} \lambda_{ij}S_iS_j \right) + {\cal L}_{\rm dark}(S_i,\psi_i,...).
\end{equation}
Here $A$ and $\lambda$ are a set of real couplings, while ${\cal L}_{\rm dark}(S_i,\psi_i,...)$ is the most general dark sector Lagrangian. 

Since we assume that at least two of the dark fields here are light, this Lagrangian induces new phenomena, both at the EW scale ({\em e.g.} novel Higgs decay channels), and at low energy, with exotic decays of $K$ and $B$ mesons. Taking $H$ field as a combination of the vacuum expectation value (VEV) and the physical Higgs field $h$, we retain a linear term in $h$: $H^\dagger H = vh +...$.
Treating all couplings in (\ref{Hportal1}) perturbatively, we can integrate out $h$ and obtain the effective Lagrangian that governs $s-d$ and $b-s$ transitions, 
\begin{equation}
    {\cal L} = \frac{3 y_t^2}{32 \pi^2 m_h^2}(V^*_{ts}V_{td} m_s \overline{s_R} d_L + h.c.)
    \left( \sum_{i} A_i S_i + \sum_{ij} \lambda_{ij}S_iS_j \right),
\end{equation}
with the similar result for the $b$ quarks upon taking $s$ to $b$. The Yukawa coupling of the top quark is defined in the standard way, $y_t= \sqrt{2}m_t/v$. 
This Lagrangian was used in many studies of $K$ and $B$ decays to new light states. For example, if one linear term dominates the sum, then one should 
expect $K\to \pi S$ two-body decays, both for charged and neutral kaons (provided of course that this is kinematically allowed).

Since we are primarily interested in the $K_L$ decays to pairs of $X$ particles, we will consider the following three scenarios:
\emph{i)} $K_L$ decays to $S_1S_2$ pair
(or similarly, to a pair of identical scalars, $S_2S_2$);
\emph{ii)} $K_L$ decays to a pair of dark fermions $X$ via intermediate scalar $S_3$; \emph{iii)} $K_L$ decays to a pair of dark scalars $S_1S_2$ via intermediate scalar $S_3$. The relevant set of couplings in each case is
\begin{align}
  &i)~~   {\cal L} \supset \lambda_{12} v \, h S_1 S_2,\\
  &ii)~~  {\cal L} \supset A_3 v \, h S_3 + y_X S_3 \overline{\psi} \psi,\\
  &iii)~~ {\cal L} \supset A_3 v\, h S_3 + B S_1S_2S_3.\label{eq:scalar-iii}
\end{align}
For all of the three cases, the amplitudes of the $K_L$ decays, as well as the on-shell Higgs boson decays to $X$ states, are given by:
\begin{align}
  &i)~~  {\cal M}_{K_L\to S_1S_2} = b \times  {\rm Im}(V^*_{ts}V_{td})F_Km_K^2 \times  \lambda_{12};&& {\cal M}_{h\to S_1S_2} = \lambda_{12} v,\\
  &ii)~~ {\cal M}_{K_L\to \psi\overline{\psi}} = b \times  {\rm Im}(V^*_{ts}V_{td})F_Km_K^2 \times  \frac{y_XA_3  m_K}{m_{S_3}^2-m_K^2};&&{\cal M}_{h\to \psi\overline{\psi}} = \frac{y_XA_3  m_h v}{m_{S_3}^2-m_h^2},\\
  &iii)~~ {\cal M}_{K_L\to S_1S_2} = b \times  {\rm Im}(V^*_{ts}V_{td})F_Km_K^2 \times  \frac{A_3B}{m_{S_3}^2-m_K^2};&& {\cal M}_{h\to S_1S_2}
  = \frac{BA_3  v}{m_{S_3}^2-m_h^2},
\end{align}
where $b = 3y_t^2/(32\pi^2m_h^2)$, $F_K\simeq 117$\,MeV, and ${\rm Im}(V^*_{ts}V_{td})\simeq 1.4\times 10^{-4}$. 

An additional decay channel of $h$ to dark particles provides strong restriction on parameter space for the model \cite{Sirunyan:2018owy,Aaboud:2019rtt}, and limits the maximal branching of $K_L \to X$ states
that one can achieve in these decays. The mass of intermediate scalar $S_3$ is a free parameter, and it turns out that it is advantageous to take $m_{S_3}$ in the intermediate range, $m_K\ll m_{S_3}\ll m_h$. With this choice we can now predict 
maximum decay rate mediated by the SM Higgs portal,
\begin{eqnarray}
   &&i)~~  {\rm BR}(K_L\to S_1S_2) = 8\times 10^{-13}\times  \frac{\Gamma_{h\to S_1S_2}/\Gamma_h^{\rm SM}}{0.1}\\
    &&ii)~~  {\rm BR}(K_L\to \psi\overline{\psi}) = 3\times 10^{-13} \times 
    \frac{\Gamma_{h\to \psi\overline{\psi}}/\Gamma_h^{\rm SM}}{0.1}\left(\frac{\rm 10\,GeV}{m_{S_3}}\right)^4 \\
     &&iii)~~  {\rm BR}(K_L\to S_1S_2) = 2\times 10^{-8} 
   \times 
    \frac{\Gamma_{h\to S_1 S_2}/\Gamma_h^{\rm SM}}{0.1} \left(\frac{\rm 10\,GeV}{m_{S_3}}\right)^4 
\end{eqnarray}
Notice that all constants $\lambda_{12}$, $y_x$, $A_3$ and $B$ entering ``dark currents" are hidden inside Higgs boson decay width to dark states. $\Gamma_h^{\rm SM}$
stands here for the SM width of the Higgs boson, $\Gamma_h^{\rm SM}\simeq 4$\,MeV. The phase space suppression for the $K_L$ decay was neglected here, effectively corresponding to 
$m_{1,2}\ll m_K$ limit. 

Examining these equations, we see that the absence of large modifications of Higgs boson decay signals due to its decay channel to dark states imposes strong restrictions on scenarios with $K_L$ decays. In particular, we find that the scenario 
\emph{i)} with the decay to two scalars via the quartic interaction with the Higgs
and case \emph{ii)} with the decay to a fermion pair via an intermediate scalar $S_3$
do not lead to any interesting signatures in $K_L$ decay. Scenario \emph{iii)}, that 
has $K_L \to S_3^* \to S_1S_2$ decay, can indeed have a large decay rate.
An extra propagator of $S_3$ leads to an enhancement of order $(m_h/m_{S_3})^4$
compared to the case \emph{i)}, which results in a much larger maximum branching ratio of $K_L$ in case \emph{iii)}.

Having identified that the model $iii)$, $K_L \to S_3^* \to S_1S_2$, can give sizeable contribution to the $K_L$ branching ratio, we would like to discuss possible observational signatures, related to instability of $S_2$ (we assume now that 
$m_2>m_1$). Within the general Higgs mediation Lagrangian (\ref{Hportal1}), 
both $S_2\to S_1 \gamma \gamma$ and $S_2\to \gamma \gamma$ are possible, 
in principle with the mixing with the Higgs field and effective
$h\gamma\gamma$ vertex. However, it is easy to see that once the constraints 
on the Higgs coupling are implemented, such decays cannot occur 
reasonably close to the production point, as the characteristic $c\tau$ will be 
way beyond 10 km distance scale. Therefore, one can expect interesting signatures in the $K_L\to X$ states decay to appear only at the expense of enlarging (\ref{Hportal1}). We give two examples of such additional terms that lead to $S_2\to S_1\pi^0$ and $S_2 \to \gamma\gamma$ decays. 

\subsubsection{\(\pi^0\) production in two Higgs doublet models}
Introduction of the second Higgs doublet, and the corresponding pseudoscalar 
Higgs field $A$ can facilitate $S_2\to S_1\pi^0$ and decay. Let us denote the extra field as $\Phi$, and keep exactly the same SM charge assignment for $\Phi$ as for $H$. 
To keep the model $SM$-like, we assume that the VEV of $\Phi$ is small, while masses of four extra scalars associated with $\Phi$ to be considerably heavier than the SM Higgs,
\begin{equation}
    \langle \Phi \rangle \ll v;~~ m_{H,A,H^\pm}\gg m_h.
\end{equation}
We will assume that $\Phi$ couples to $\overline{Q} D$ quark bilinears 
with the Yukawa matrix $Y_d^\Phi$ $\propto Y_d$ in order to 
remove extra Higgs-mediated FCNC effects, and preserve MFV. 
Relation $Y_d^\Phi$ $\propto Y_d$ implies proportionality of matrices, but we note that individual values of the Yukawa couplings can be much larger for 
$\Phi$ than for the SM, $y_d^\Phi\gg y^{\rm SM}_d$, which we will assume to be the case. 
Also, the dominance of the down-type Yukawa couplings of $\Phi$ justifies neglecting 
$t-W$ loop-induced FCNCs.
Extra terms in Lagrangian that will introduce 
$S_2\to S_1\pi^0$ decay are chosen to be 
\begin{equation}
\label{Phi0}
    {\cal L}_{\Phi} = \lambda^\Phi i(\Phi^\dagger H - H^\dagger \Phi)S_1S_2
    + y_d^\Phi (\overline{d_L} d_R \Phi^0 + \overline{d_R} d_L (\Phi^0)^*),
\end{equation}
where $\Phi^0 =2^{-1/2}(H+iA)$ is the neutral component of $\Phi$.
This Lagrangian leads to the following decay amplitude
\begin{eqnarray}
   {\cal L}_{eff} = S_2S_1\pi^0 \times m_{eff};\qquad{\cal M}_{S_2\to S_1 \pi^0} = m_{eff}= y_d^\Phi \lambda^\Phi \times 
   \frac{v\langle \overline{q}q \rangle}{\sqrt{2}F_\pi m_A^2},
   \label{meff}
\end{eqnarray}
where $F_\pi = 92$\,MeV and $\langle \overline{q}q \rangle$ is the light quark vacuum condensate value, which we take to be $\sim (250\,{\rm MeV})^3$. Taking 
$m_A$ to be commensurate with 1 TeV, and the product of coupling constants on the 
order $10^{-3}$, one arrives to a reasonably fast decay rate,
\begin{equation}
\label{S2toPi0}
    \Gamma_{S_2\to S_1 \pi^0}= \frac{1}{c\tau_{S_2}} \simeq \frac{1}{3.5\,{\rm m}}\times
    \left(\frac{y_d^\Phi \lambda^\Phi}{10^{-3}}\right)^2 \left(\frac{{\rm TeV}}{m_A}\right)^4 \frac{300\,{\rm MeV}}{m_2}\times \lambda^{1/2}(1,y_1^2,y_\pi^2),
\end{equation}
where here and throughout we define $y_a=m_a/m_2$, and make use of the K\"all\'en function, $\lambda(a,b,c) = (a-b-c)^2-4bc$.
This formula assumes that the decay of $S_2$ is dominated by this mode, while other 
decay channels may exist as well ({\em e.g.} $S_2 \to 2S_1$ and/or $S_2\to 3S_1$), in which case $\tau_{S_2}$ may turn out 
to be considerably shorter.
One can see that even for moderately small values of the Yukawa couplings and masses 
for an extra pseudoscalar Higgs, the decay is relatively prompt. In the next section, 
we will address in more detail the expected signature of $K_L \to S_1S_2$ at the KOTO experiment. 

We also note that the existence of $S_1-S_2-\pi^0$ vertex implies a corresponding $\eta$ vertex and its decay to $S_1S_2$ with a BR given by
\begin{equation}
   {\rm BR}(\eta\to S_1S_2) =  1.2\times 10^{-8} \left(\frac{y_d^\Phi \lambda^\Phi}{10^{-3}}\right)^2 \left(\frac{{\rm TeV}}{m_A}\right)^4 
   \times \lambda^{1/2}(1,z_1^2,z_2^2),
\end{equation}
where here and throughout the text, we define $z_a = m_a/m_\eta$. By itself, such a small BR does not seem to pose any constraints
from studies of $\eta$ meson decay. However, this is a very large branching for the beam dump production of $\eta$ with subsequent decay of $S_2$ inside a detector. A boost on the order $O(100)$ will easily take (\ref{S2toPi0}) to hundred meters length scale, which is enough to put this scenario in trouble unless the decay length is shortened, either due to larger values of $ y_d^\Phi \lambda^\Phi m_A^{-2}$, or due to {\em additional} decay channels, such as $S_2 \to 2S_1$ or $3S_1$. 

\subsubsection{A \(\pi^0\) impostor from effective coupling to photons} 
A separate interesting possibility emerges when $S_2$ is intrinsically unstable. If, for example, 
$S_2$ couples to the vector-like with respect to the SM heavy fermions $\Psi$ via 
$\lambda_\Psi \overline{\Psi} i \gamma_5 \Psi S_2$, then a loop of $\Psi$ will generate 
an effective couplings to photons. Assigning the charge of $\Psi$ to the SM photons to be 1, we obtain the following effective coupling to photons, 
\begin{equation}
    {\cal L}_{eff} = \frac{\alpha\lambda_\Psi}{4\pi m_\Psi} S_2 F_{\mu\nu} \tilde F_{\mu\nu},
\end{equation}
and the decay rate $S_2\to \gamma\gamma$,
\begin{equation}
    \Gamma_{S_2\to \gamma\gamma}= \frac{1}{c\tau_{S_2}} \simeq \frac{1}{2\,{\rm m}}\times\left(\frac{{\rm TeV}}{(m_\Psi/\lambda_\Psi)}\right)^2 \left(\frac{m_2}{m_\pi}\right)^3.
\end{equation}
As in the previous case, this decay is relatively prompt for a TeV-scale vector-like mediator.

To conclude this sub-section, the SM Higgs portal provides a very natural 
realization of the MFV $s-d$ (and $b-s$) transition. The non-minimal version of such portal with multiple scalar fields, can lead to a sizeable, $O(10^{-8})$, branching 
ratios of $K_L$ mesons to pairs of such scalars. Subsequent decay of one these scalars to $\gamma \gamma$, or to $S_1\pi^0$ will lead to observable signatures 
mimicking $K_L \to \pi \overline{\nu}\nu$ decays.

To conclude this subsection, the Higgs mediation is capable of inducing 
sizeable $K_L \to X_1X_2$ decays with subsequent decay of one of these 
state to $\pi^0$ or photons. We find, however, that the model-building 
options in this case are non-minimal, and different 
couplings must be responsible for $K_L$ decay and for further fragmentation of dark states. We summarize our findings in Table \ref{tab:model_summary}.

\subsection{UV completion via \(Z^\prime\) portal}\label{sec:zportal}

In this subsection we discuss UV completion of vector operator in (\ref{operators}). The completion necessitates the introduction of new vector boson(s), 
that we will call $Z'$. There is vast amount of literature on the so-called 
``dark photon", or a new particle $A'$ coupled to the SM via a kinetic mixing portal, 
$(\epsilon/2)F_{\mu\nu}F'_{\mu\nu}$. Since $A'$ is massive, it can be coupled to non-conserved dark currents, such as $X_1 \partial_\mu X_2 - X_2 \partial_\mu X_1$,
if $m_1\neq m_2$. (This current, in turn, cannot be fundamental, and may be the result of a mass splitting induced by some dark Higgs condensation mechanism.)
However, this does not lead to enhanced $K_L\to X_1X_2$ decay. The reason is that 
the photon penguin diagram induces the following effective operator: $\overline{s_L} \gamma^\mu d_L \times \partial^\nu F_{\mu\nu}$. When the quark current 
is replaced with the momentum of $K_L$ this leads to the vanishing of the amplitude
on account of $\partial^\mu\partial^\nu F_{\mu\nu}=0$. In addition, the diagram with kinetic mixing of the dark photon with the SM $Z$-boson is proportional to $\epsilon \,m_{K}^2/m_Z^2$, and does not lead to any enhancement in $K_L$ decays.

Thus, we turn our attention to $Z^\prime$ models coupled to non-conserved currents. 
The example of photon penguin teaches us that the coupling of $Z'$ to quark current 
must be of the form $\overline{s_L} \gamma_\mu d_L Z^\prime_{\mu}$. The general analysis of 
these options was performed in {\em e.g.} Ref. \cite{Dror:2017nsg}. 
Here, for concreteness, we choose a model based on 
mass-mixing between the SM $Z$ and $Z^\prime$ bosons. 

If the mass mixing between the new vector boson $Z'$ and the SM $Z$ boson~\cite{Babu:1997st,Davoudiasl:2012ag,Davoudiasl:2012qa,Davoudiasl:2013aya} is the main source of coupling between the two sectors, then the resulting flavor physics is automatically complying with the MFV expectations. 

In generality, the relevant terms may be written as  
\begin{equation}\label{eq:XEFTmassmixing}
    \mathcal{L}  \supset \frac{1}{2}m_{Z^0}^2Z^0_\mu Z^{0\mu}-\Delta^2 Z^0_\mu X^\mu+\frac{1}{2}m_X^2X_\mu X^\mu,
\end{equation}
where $Z^0_\mu$ denotes the SM-like $Z$ boson in the mass basis with respect to the photon and $Z$ fields (not the mass eigenstate incorporating the $\Delta$ term), namely, $m_{Z^0}=g_Z v/2$ with $g_Z=g/c_W$ and $c_W$ being the cosine of the weak angle.
Then, in the mass eigenbasis for neutral gauge bosons ($A_\mu$, $Z_\mu$, $Z^\prime_\mu$), we have $m_Z^2\simeq m_{Z^0}^2$ and $m_{Z'}^2\simeq m_X^2-\Delta^4/m_Z^2$.
The mass mixing parameter $\varepsilon_Z$ is then defined as $\varepsilon_Z=\Delta^2/m_Z^2$.
For a gauge invariant description, we resort to the SM+$X$ effective field theory description from Ref.~\cite{Dror:2018wfl}, where the effects of the following mass mixing operator were discussed,
\begin{equation}\label{eq:XEFTmassmixing2}
    \mathcal{L} \supset g_X X^\mu i \left( C_1 H^\dagger \overleftrightarrow{D_\mu}_{Z} H \right) \xrightarrow[]{\text{EWSB}}
 \varepsilon_Z m_Z^2 X_\mu Z^\mu, 
\end{equation}
where $C_1$ represents a coefficient that reflects the UV completion of this operator.
Such an operator has been extensively discussed in the literature~\cite{Babu:1997st,Davoudiasl:2012ag,Davoudiasl:2012qa,Davoudiasl:2013aya}, and in the mass eigensbasis for neutral gauge bosons, leads to the following couplings for the new massive $Z^{\prime\mu}$ and the SM-like $Z^\mu$:
\begin{equation}\label{eq:XEFTmassmixing3}
    -\mathcal{L} \supset \varepsilon_Z \left(\frac{g}{2 c_W} J_\mu^{\rm NC} + e J_\mu^{\rm EM}\right)\,Z^{\prime\mu}  +  \frac{g}{2 c_W} J_\mu^{\rm NC}\,Z^\mu,
\end{equation}
where $J_\mu^{\rm NC}$ and $J_\mu^{\rm EM}$ are the standard neutral and electromagnetic currents in the SM. It is the coupling of $Z'$ to $J^{\rm NC}_\mu$
that would lead to operator (\ref{operators}).

The quark-$W$ loop induces the coupling (see {\em e.g.} Ref.~\cite{Dror:2018wfl})
\begin{equation}
   \mathcal{L} \supset 
   g_{sdX} \,X_\mu\, \left( \overline{s} \gamma^\mu P_L d \right)\,+\, \text{h.c.}, \quad \text{where}\quad g_{sdX}  \simeq \frac{g^3 \, \varepsilon_Z}{32 \pi^2 c_W} \sum_{i} V_{i s}V^*_{i d} f\left(\frac{m_{i}^2}{M_W^2}\right),
\end{equation}
where the real part of this coupling will induce $K_L\to X_1X_2$ decay.
It is important to recognize that the mass mixing of $Z$ and $Z^\prime$ needs further UV completion, and as a consequence loop function $f(x)$ is logarithmically enhanced if the scale of such completion is taken to be large. For the current purposes, we will take the form $f(x) = -(x/4)\log{m_{\rm UV}^2/M_W^2}$, where the logarithm receives a cut-off by the mass of particles that 
"resolve" $ \varepsilon_Z$~\cite{Dror:2018wfl}, and we take this scale $m_{\rm UV} = 500$ GeV for concreteness. With this choice, we find that the $s-d$ transitions, always suppressed by either $V_{ts} V_{td}^*$ or $m_{u,c}^2/M_W^2$, depend on a small quantity $g_{sdX}/\varepsilon_Z \simeq (1.45 + i\,6.15) \times 10^{-6}$. Among possible UV completions, the most straightforward one is the two-Higgs doublet model, where the additional Higgs boson field, charged under $U(1)_X$, does not couple to quarks. Its VEV being much smaller than the SM-like doublet leads to suppressed effects in EW observables such as the $Z$ and Higgs boson decay. 
Still there is a possibility of substantial cancellation between the new physics logarithmic piece and the SM $b-s-Z$ amplitude \cite{Dror:2018wfl}, so that $f(x)$ can further deviate from the value adopted here. 

For concreteness, we take the dark current in the following form, 
\begin{equation}\label{eq:Lscalarpair}
    \mathcal{L}_S \supset g_X Z'_\mu J_S^\mu = g_X Z'_\mu (S_2\partial^\mu S_1 - S_1\partial^\mu S_2), 
\end{equation}
where $S_{1,2}$ are real scalar fields. The decay to two different SM-singlet fermions can occur via
\begin{equation}\label{eq:Lfermionpair}
\mathcal{L}_{\psi} \supset  g_X Z'_\mu J_\psi^\mu = g_X \, Z^\prime_\mu  \left(c_V \overline{\psi_2} \gamma^\mu \psi_1  +  c_A
\overline{\psi_2} \gamma^\mu \gamma_5 \psi_1  + \mathrm{h.c.}\right).
\end{equation}
The decay to particle and its anti-particle, $K_L\to \overline{\psi}\psi$ is possible but can occur only due to axial-vector coupling to $Z^\prime$.

If $\psi_{1}$ and $\psi_2$ are Majorana fermions, then $c_V$ ($c_A$) is purely imaginary (real) with $g_X$ real. In all the vector portal cases, it is instructive to define a phenomenological contact-interaction coupling as follows
\begin{equation}
    \frac{G_X}{\sqrt{2}} = \frac{\varepsilon_Z g\, g_X}{4 c_W m_{Z^\prime}^2},
\end{equation}
where $g$ is the weak coupling, and $G_X$ is to be compared with the SM Fermi constant $G_F$. 

With this input, one can predict $K_L\to S_1S_2$ amplitude and the branching ratio, 
\begin{align}
   {\cal M }_{K_L \to S_1S_2} &= \frac{{\rm Re}(g_{sdX})g_XF_K}{m_{Z'}^2}\times (m_2^2-m_1^2)\\
   {\rm BR}(K_L\to S_1S_2) &= 1\times10^{-8}\times  \left(\frac{G_X}{G_F}\right)^2\left[\frac{m_2^2-m_1^2}{(300\,{\rm MeV})^2}\right]^2 \lambda^{1/2}(1,r_1^2,r_2^2),
\end{align}
where here and throughout we define $r_a=m_a/m_{K_L}$
While this branching may appear relatively small, we note that 
$\varepsilon_Zg_X/m_{Z'}^{2}$ can be somewhat larger than $G_F$ owing to the 
possibility of having $m_{Z'} \ll m_Z$. Unlike the case of the Higgs mediation, 
where only scalar final states in the dark $K_L$ decays could give sizeable rates,
the $Z^\prime$ mediation can also be realized with final state fermions. 

For Dirac fermions, we compute the total BR into both $\psi_1\overline{\psi_2}$ and $\psi_2\overline{\psi_1}$, and find
\begin{equation}
     \text{BR} \left( K_L \to \psi_1 \psi_2\right)_{\rm D} =  3\times10^{-7}\times \left(\frac{G_X}{G_F}\right)^2\left[ |c_V|^2 \Delta r^2 (1-r^2) + |c_A|^2 r^2 (1-\Delta r^2) \right] \lambda^{\frac{1}{2}} \left( 1, r_1^2, r_2^2\right).
\end{equation}
where $\Delta r=r_2-r_1$, $r=r_1+r_2$. The computation is analogous for Majorana fermions, and shows that the BR is twice as large in that case, $\text{BR} \left( K_L \to \psi_1 \psi_2\right)_{\rm M} =  2 \,\text{BR} \left( K_L \to \psi_1 \psi_2\right)_{\rm D}$, in agreement with the analogous calculation of $K_L\to\nu\overline{\nu}$ in Ref.~\cite{Marciano:1996wy}.

Stepping ahead, the most interesting consequences for the $K_L$ and 
$B$ meson decays occur for the range of $G_X$ comparable to $G_F$. The least constrained possibility from direct collider searches is when the mass of $Z^\prime$ is sub-electroweak scale, but above the scale of direct production at $B$ factories ($\sim 10$\,GeV range). At the same time, we would like to keep 
$g_X$ sizeable, $O(1)$, and $\varepsilon_Z$ to be rather small, in the  $10^{-3}-10^{-2}$ range. One could question if this choice of parameters can be 
realized in models that provide UV completion to $ \varepsilon_Z$ parameter. Because $g_X$  is large, achieving small $\varepsilon_Z$ at tree level by condensing an additional Higgs field $H_X$ 
that carries both SM and $U(1)_X$ charges can be difficult, unless 
$H_X$ charge under $U(1)_X$ is very small. Of course, a minimal solution is to write $M_{2}^2 |H_X|^2$ as a positive mass term, and instead add a new SM-singlet complex scalar $\phi$ that breaks the $U(1)_X$ by a new VEV $v_\phi$. In that case, a tadpole term $\mu (H^\dagger H_X) \phi$ induces a VEV for $H_X$ of $v_X \simeq (\mu v_\phi v/ M_{2}^2) /2\sqrt{2}$, and it is easy to show that $m_{Z^\prime} \simeq g_X v_\phi$ remains large, while $\varepsilon_Z \simeq (2 g_X/g \,c_W) (v_X/v)^2$ can be made very small. By taking $M_2 \gg v^2$, all scalars associated with $H_X$ escape detection for being very heavy, while a new dark scalar $\Re(\phi)$ remains lighter than the Higgs, implying that the mixing $\lambda_{\phi H} |\phi|^2|H|^2$ ought to be small. Yet another way of generating mass mixing would be through loops of particles, beyond SM fermions and/or bosons, that are charged under both SM and $U(1)_X$. We need the mass mixing, rather than 
kinetic mixing of $Z\!-\!Z^\prime$, and therefore the mass of particles in the loop $m_P$
{\em must} receive contributions both from the SM Higgs VEV $v$ and the VEV 
of $H_X$, $v_X$. In other words, schematically, $m_P = m_0 + c_1 v +c_2 v_X$. 
Then one loop effect will generate $\varepsilon_Z$ as effectively 
dimension-6 operator that decouples as $(v_X)^2/m_0^2$ in the large $m_0$ limit.
Therefore, having a small $ \varepsilon_Z$ and large $g_X$ are not incompatible.

\subsubsection{{\(\pi^0\)} production from \(Z^\prime\!-\!Z\) mixing }\label{sec:zportal_pi0}

An interesting feature of the $Z'$ mediation is the possibility of $X_2\to X_1\pi^0$ 
decay mediated by $Z^\prime\!-\!Z$ mixing, as the same coupling also appears in $K_L$ decays. 
Calculating the matrix element, we get 
\begin{equation}
    {\cal M}_{S_2\to S_1 \pi^0} =
\sqrt{2} G_XF_\pi(m_2^2-m_1^2),
\end{equation}
which leads to the decay rate 
\begin{equation}
    \Gamma_{S_2\to S_1\pi^0}= \frac{1}{c\tau_{S_2}} =\frac{1}{16\,{\rm cm}}\times \left(\frac{G_X}{G_F}\right)^2 \left(\frac{m_2}{300 \text{ MeV}}\right)^3 \left(1 - y_1^2\right)^2 \, \lambda^{1/2}(1,y_1^2,y_\pi^2),
    \end{equation}
The effectively weak-strength interaction will induce the decay of $S_2$ at distances 
comparable to geometry of experiments with $K_L$. 
We would like to note that the value of the $S_2\to S_1\pi^0$
decay amplitude for the choice of $G_X=G_F$ and $m_2^2-m_1^2 = (300\,{\rm MeV})^2$ is approximately ${\cal M}_{S_2\to S_1 \pi^0} \simeq 140$\,eV.
With the same choices, the amplitude for the $K_L$ decay is $\sim 2.5\times 10^{-3}\,\rm eV$.

We also note that the amplitude of $\eta\to S_1S_2$ decay and the corresponding 
branching ratio is given by
\begin{equation}
    {\cal M}_{\eta\to S_1 S_2} =
\sqrt{2/3} G_XF_\pi(m_2^2-m_1^2);~~BR(\eta\to S_1S_2) = 2\times 10^{-9}
\left(\frac{G_X}{G_F}\right)^2(z_2^2-z_1^2)^2\lambda^{1/2}(1,z_1^2,z_2^2),
\end{equation}
Again, this is a substantial rate for the beam dump experiments where $\eta$ mesons are produced, if the lifetime 
of $S_2$ is in the right range for a decay at a distant detector. 

In the Dirac fermion case, we find the decay length parameter to be
\begin{align}
        & \Gamma_{ \psi_2\to\psi_1 \pi^0}^{\rm D} = \frac{1}{c\tau_{\psi_2}} =\frac{1}{16 \text{ cm}}\times \left(\frac{G_X}{G_F}\right)^2\left(\frac{m_2}{300 \text{ MeV}}\right)^3\left[|c_V|^2 F(-y_\pi,-y_1) + |c_A|^2 F(y_\pi,y_1)\right] \lambda^{1/2}(1,y_1^2,y_\pi^2),
\end{align}
where $F(y_\pi,y_1) = (1+y_1)^2((1-y_1)^2-y_\pi^2)$. For Majorana fermions, the previous decay rate are larger by a factor of two. Eta  meson decays to Dirac $\psi_1 \psi_2$ pairs, accounting for both $\overline{\psi_1} \psi_2$ and $\psi_1 \overline{\psi_2}$, are given by
\begin{equation}
    \Gamma_{\eta \to \psi_1\psi_2 }^{\rm D} = \frac{F_\pi^2 G_X^2 m_\eta^3}{6 \pi}\left[|c_V|^2(z_1-z_2)^2(1-(z_1+z_2)^2)+|c_A|^2(z_1+z_2)^2(1-(z_1-z_2)^2)\right] \lambda^{1/2}(1,z_1^2,z_2^2),
\end{equation}
which results in branching ratio $\sim 8\times 10^{-9}$ for the fiducial choice of parameters. 

\subsubsection{Dipole portal}

Another possibility discussed in this paper is the fermionic state $\psi_2$ decaying to the $\psi_1\gamma$ due to the dipole operator. 
A pair of $\psi_2$ fermions can be produced via the $Z'$ axial-vector current. If, in addition, there is an effective dimension five coupling
to $X_1$,
\begin{equation}
    {\cal L} = \frac{\mu}{2} \,\overline{\psi_1} \sigma_{\mu\nu}\psi_2\,F^{\mu\nu} + h.c., 
\end{equation}
then 
the decay rate is given by \cite{Lee:2014koa} (see also \cite{Pospelov:2013nea})
\begin{equation}
 \Gamma_{\psi_2\to \psi_1\gamma }^{\rm D} = \frac{\mu^2m_2^3}{8\pi}\left(1-\frac{m_1^2}{m_2^2}\right)^3  =\frac{1}{c\tau_{\psi_2}} =
  \frac{1}{5\,{\rm cm}} \left(\frac{\mu}{(100\,\rm TeV)^{-1}}\right)^2 
    \left[\frac{m_2}{100\,{\rm MeV}}\right]^3\left(1-y_1^2\right)^3.
\end{equation}
Thus, the decay of $\psi_2$ can be rather prompt for a very small values of 
the effective dipole moment. The UV completion of $\mu$ itself would require introduction of heavy states charged under electromagnetism and coupled to 
$\psi_{1,2}$. 

We conclude by emphasizing that the $Z^\prime$ portal may provide a flexible way of arranging $K_L$ decays to pairs of dark states, that subsequently fragment to give two photons, either individually, or via $\pi^0$ production. In the latter case, a single combination of couplings controls $K_L$ and $X_2$ decay. The summary of these models is included in Table \ref{tab:model_summary}.

\subsection{\(K_L\) decays via virtual \(\pi^0,\, \eta\)}\label{sec:virtualpi0}

So far, we have considered FCNC type of $s\to d$ transition associated either with a $Z'$ or Higgs mediation, and in both cases $W-t$ loop and 
more generically short-distance contributions play the most important 
role. In this subsection, we consider regular $\Delta S = 1$ quark flavor of the SM, the long-distance part of it, paired with the 
{\em flavor-diagonal} coupling of quarks to the two dark scalars. 

More specifically, we concentrate on the following process:
\begin{equation}
    K_L\to {\rm virtual~}\pi^0~{\rm or}~ \eta \to S_1S_2 \to 2S_1+\pi^0.
    \label{virtualpi}
\end{equation}
This is an attractive phenomenological possibility, as the same effective ${\rm meson}-S_1-S_2$ coupling governs both decays, $K_L\to S_1S_2$ and $S_2\to S_1 \pi^0$. 

As is well appreciated in the SM flavor literature, 
the mixing of $K_L$ and light non-strange pseudoscalar mesons is induced by the
long-distance part of the $\Delta S =1$ (see {\em e.g.} \cite{DAmbrosio:1994fgc} and references therein). We shall assume the leading order $SU(3)$ chiral perturbation theory treatment and neglect the contribution 
of $\eta'$ to find the following transition amplitudes:
\begin{eqnarray}
{\cal M}_{K_L-\pi^0} = -0.07 \,{\rm MeV}^2;\qquad{\cal M}_{K_L-\eta} = \frac{1}{\sqrt{3}}{\cal M}_{K_L-\pi^0} = -0.04 \,{\rm MeV}^2
\end{eqnarray}
The mixing matrix element is extracted with the use of 
soft-pion theorem from experimental data on $K_S\to \pi\pi$ \cite{Ma:1981eg}. 

Next, we shall assume a pseudoscalar Higgs mediated, flavor-conserving couplings between
light quarks bilinears $\overline{q} i\gamma_5 q$ and two light scalars, $S_1S_2$. We follow the model given in \refeqs{Phi0}{meff}:
\begin{equation}
\label{dim5}
    {\cal L} = \frac{v\lambda^\Phi}{\sqrt{2}m_A^2}(y_d^\Phi \overline{d} i\gamma_5d
    + y_s^\Phi \overline{s} i\gamma_5s)S_1S_2.
\end{equation}
The MFV prescription tells us that $y^\Phi_s/y^\Phi_d = y^{\rm SM}_s/y^{\rm SM}_d \simeq 19$, where we used $m_s/m_d$ determination of \refref{Leutwyler:1996qg}.
Translating (\ref{dim5}) into couplings to pseudoscalar mesons, and introducing 
$m_{eff}$ parameter as before, \refeq{meff}, we get 
\begin{equation}\label{eq:virtualpi0decay}
    {\cal L} = m_{eff} S_1S_2 \left(\pi^0 +
    \eta\times \frac{2}{\sqrt{3}}\times \frac{y^{\rm SM}_s}{y^{\rm SM}_d}\right)
    \simeq m_{eff} S_1S_2 \left(\pi^0+22.\times\eta\right).
\end{equation}
We then can derive a decay amplitude in terms of $m_{eff}$:
\begin{equation}
\label{resAmpl}
    {\cal M}_{K_L\to S_1S_2} = m_{eff} \left(\frac{{\cal M}_{K_L-\pi^0}}{m_K^2-m_\pi^2}
    + \frac{{\cal M}_{K_L-\eta}\times 22.}{m_K^2-m_\eta^2}\right)
    \simeq 1.7\times 10^{-5}\times m_{eff} .
\end{equation}
The contribution of $\eta$ to (\ref{resAmpl}) is much enhanced due to a smaller denominator and $m_s/m_d$ in the numerator. These expressions lead to the 
following relevant quantities,
\begin{align}
    {\rm BR} (K_L\to S_1S_2) &= 9\times 10^{-9} \times \left( 
    \frac{m_{eff}}{\rm 100\,eV} \right)^2\lambda^{1/2}(1,r_1^2,r_2^2),
    \\
    {\rm BR} (\eta\to S_1S_2) &=1.3\times 10^{-7} \left( \frac{m_{eff}}{\rm 100\,eV} \right)^2\lambda^{1/2}(1,z_1^2,z_2^2),
    \\
    \label{eq:S2pi0decay}
    \Gamma_{S_2\to S_1 \pi^0} &= \frac{1}{c\tau_{S_2}} \simeq \frac{1}{30\,{\rm cm}}\times
    \left( \frac{m_{eff}}{\rm 100\,eV} \right)^2 \left(\frac{300\,{\rm MeV}}{m_2}\right) \lambda^{1/2}(1,y_1^2,y_\pi^2).
\end{align}
Thus we observe that a meson$-S_1-S_2$ vertex controlled by 
$m_{eff}$ in the range of 100\,eV simultaneously lead to 
sizeable decay rates of $K_L$, and to relatively prompt decays of $S_2$.
How realistic is to expect $m_{eff}$ to be in this range? As seen before, 
the choice of the Yukawa couplings in the $10^{-3}-10^{-2}$ range, and 
$m_{A}$ close to a TeV benchmark gives $m_{eff}$ in this ballpark. 

There are several limitations to this scenario. One comes from the Yukawa coupling of the $b$-quark, $y^\Phi_b = (m_b/m_d)y^\Phi_d$, which 
cannot be chosen much above one in order to preserve perturbativity. Together with the LHC constraints on Higgs fields from $\Phi$ doublet this sets the maximum of $m_{eff}$ parameter, that cannot be taken much above the keV mark. 
We also note that this model is significantly constrained by the combination of beam dump experiments
and $K_L$ decays:
compliance with CHARM bounds \cite{Bergsma:1985qz} (see their recent evaluation in \cite{Darme:2020ral}), 
restricts lifetimes of $S_2$, and requires $m_{eff} > 100$\,eV. At the same time, KOTO results generally 
require $m_{eff} < 100$\,eV so that 
$K_L\to S_1S_2$ be comparable 
with the current bounds.  
To make this model fit all the constraints easier one could introduce another parameter, {\em e.g.}
$\lambda_{12} S_1^3S_2$ coupling, that would shorten the lifetime of $S_2$ provided that $m_2>3m_1$
and reduce $K_L\to S_1S_2\to \pi^02S_1$ rate. Then, for example, $m_{eff} = 200$\,eV and BR$(S_2\to S_1\pi^0)= 0.1$ would not be probed by the beam dump experiments, and would predict sensitivity $K_L\to \pi^0\slashed{E}$ close to the edge of the current exclusion bounds.

At the same time, it is clear that one cannot construct a realistic model that would 
have large $K_L$ decay rate 
to a fermionic dark pair mediated by virtual $\pi^0$ and $\eta$. 
This is because the analogue of \refeq{dim5} will 
contain dimension six, rather than dimension five, operators. As a result, 
the branching ratio of $K_L\to \overline{X_2}X_1\,{\rm or }\,\overline{X_1}X_2$ will be 
suppressed relative to the scalar case by an additional factor that is at least as small as $ (\Lambda_{\rm hadr}/{100\,\rm GeV})^2<10^{-4}$. 

The final comment we would like to make is about a  ``minimal" phenomenological possibility away from the MFV point. If, for example, only the coupling to 
the down quark exists, $\Lambda^{-1} (\overline{d} i \gamma_5 d)S_1S_2$, then the effective Lagrangian would not have an enhanced coupling to $\eta$, ${\cal L} 
= m_{eff} S_1S_2 (\pi^0-3^{-1/2}\eta)$. 
Then, the choice of $m_{eff}=1\,{\rm keV}$ would lead to a BR$(K_L\to S_1S_2)$ at level of $2\times 10^{-9}$ and subsequent decay of $S_2$ to $S_1\pi^0$ with a $c\tau_{S_2}$ of less than 1 cm, leading to interesting predictions for current experimental facilities.
Such a model would employ $\Lambda \simeq 170$\,TeV 
and be perfectly consistent with effective field theory, but suffer from extreme sensitivity to assumed values for $(\overline{s} i \gamma_5 s)S_1S_2$ etc operators, and possibly new sources of FCNC associated with the breakdown of MFV ansatz. 

\subsection{Overview}

We have identified multiple models for pair production of new particles in $K_L$ decays and shown that several scenarios can lead to an enhancement of $K_L\to\pi^0 \slashed{E}$ rates with respect to the SM branching ratio. We can categorise our models in three distinct proposals depending on how the visible $\gamma\gamma$ signature arises after pair production. These are shown in \reffig{fig:diagrams}, and correspond to the possibilities in (\ref{pi0production}), (\ref{2gproduction}), and (\ref{dipoleproduction}). The successful proposals for enhancing $K_L$ decays are summarised in \reftab{tab:model_summary}. Due to the Lorentz structure of the vertex, only fermions can be considered in the dipole case, and only scalars may play the role of a $\pi^0$ impostor. Note that our survey does not exhaust the list of possibilities. In fact, other minimal models may arise, for instance, with heavy pseudoscalar mediators, which is expected to be analogous to our vector portal model. We also did not consider pair production of vector particles, although such a possibility is in principle allowed, for example with non-abelian dark sectors.
\newcommand{\success}{$\checkmark$}
\newcommand{\failure}{$-$}
\begin{table}[h]
    \centering{\small
    \begin{tabular}{|c|c|c|c|c|}
        \hline
        \multirow{2}{*}{Scenario}
        & \multicolumn{2}{|c|}{(A) $\pi^0$ production} & (B) dipole portal & (C) $\pi^0$ impostor
        \\\cline{2-5}
       & $X_i=S_i$&$X_i=\psi_i$&$X_i=\psi_i$& $X_i=S_i$
        \\\hline
        Vector portal &\success&\success&\success&\failure \\
        Scalar portal&\success&\failure&\failure&\success \\
        Virtual $\pi^0,\eta$ & \success&\failure&\failure&\failure \\
        \hline
    \end{tabular}}
    \caption{Summary of the scenarios proposed in \refsec{sec:models}. A checkmark indicates that a successful enhancement to $K_L\to X_1 X_2$ with respect to the SM $K_L\to\pi^0\overline{\nu}\nu$ value was possible. Dashes indicate either not studied or no enhancement was possible.}
    \label{tab:model_summary}
\end{table}

Many of the scenarios we proposed called for two distinct combinations of couplings: one combination for production and one additional coupling to facilitate the decay of the heavier dark states to a $\pi^0$-like signature. This is the case in all our scalar portal models, especially the viable model given by \refeqs{eq:scalar-iii}{Phi0}, as well as in the dipole portal decay. A notable exception to this is scenario (A), both in the vector portal model, \refeqs{eq:Lscalarpair}{eq:Lfermionpair}, as well as with virtual $\pi^0$, $\eta$ production, \refeqs{dim5}{eq:virtualpi0decay},  where the same product of coupling controls both $K_L$ and $X_2$ decays. Due to the reduced number of couplings, we will investigate scenario (A) in more detail to find the allowed parameter space in terms of $G_X$, $m_1$, and $m_2$. We still consider (B) and (C) in the general discussion of signal reconstruction without studying the existing constraints on the couplings.

\section{Case study at KOTO and \(B\) factories}
\newcommand{\pt}{|\vec{p}_\pi^{\,T}|}
\newcommand{\ptK}{|\vec{p}_K^{\,T}|}

In this section, we study the different (A) to (C) scenarios in more details in the context of $\overline{\nu}\nu$ modes in neutral $K$ meson experiments. We pay special attention to the KOTO experimental setup, as our exotic signatures mimic signal $\pi^0$ events through mis-reconstruction. Our goal is to explore limits imposed by published KOTO data on the models considered here, and,
in a more speculative vein, explore if some of these models could account for the recently reported excess events.
Due to the reduced freedom in the parameter space of the $Z^\prime$ portal model discussed in \refsec{sec:zportal} and the virtual $\pi^0$, $\eta$ scenario discussed in \refsec{sec:zportal_pi0}, we restrict our discussion to these cases. We will also compute the sensitivity of $B$ factories such as Belle II to the $Z\!-\!Z^\prime$ model for pair production.

\subsection{The KOTO experiment and analysis strategy}
The KOTO detector is located in a neutral kaon beam at J-PARC~\cite{10.1093/ptep/pts057}. The beam is $16^\circ$ degrees off-axis from the 30 GeV proton beam with a peak $K_L$ momentum of $1.4$~GeV, and provides about $4.2\times10^7$ $K_L$’s per $2\times 10^{14}$ protons on target. The detector comprises a cylindrical vessel enclosing a $\sim$3~m long decay volume followed by a $95$~cm radius electromagnetic calorimeter (ECAL) with a square beam hole of $15\times15$ cm$^2$. The inner walls of the detector as well as the front surface of the calorimeter are equipped with charge and photon vetoes to reject any events involving charged particles or additional photons. As we will see, to acutely reconstruct the $\pi^0$ momentum, it is important that the $K_L$ beam be a ``pencil beam", that is, that it be a sufficiently narrow beam. The beam at KOTO has about $8\times8$ cm$^2$ transverse area and a very small $K_L$ transverse momentum, typically below few MeV (see \refapp{app:pt}). In our discussion, we adopt the standard KOTO detector coordinates, with the Z axis defined as the central axis along the beam line, with $Z=0$ constituting the front barrel and $Z=686$~mm the surface of the ECAL.

We briefly outline the analysis strategy at KOTO in what follows (for additional details, see Refs.~\cite{Ahn:2016kja,Su:2019xfn,Nakagiri:2019yec}). The signal selection is primarily based on two reconstructed quantities: $\pt$, the transverse momentum  of the pion with respect to the $Z$ axis, and  $Z_{\rm vtx}$, the reconstructed $Z$ coordinate of the $K_L$ decay inside the detector. Neutrinos can take away a large amount of transverse momentum in $K_L\to\pi^0\nu\overline{\nu}$ decays, so requiring large $\pt$ can help reduce backgrounds. Ignoring any transverse momentum of the initial kaon beam, the maximum value of $\pt$ in a general $K_L\to\pi^0 Y_2 Y_3$ decay is given by
\begin{equation}\label{eq:3bodypt}
    \pt_{\rm true} <  \frac{1}{2 m_{K_L}} \lambda^{1/2}\left(m_{K_L}^2,m_\pi^2,(m_2+m_3)^2\right),
\end{equation}
where for $K_L\to\pi^0\nu\overline{\nu}$ ones finds $\pt \lesssim 230 \,{\rm MeV}$, while in $K_L\to\pi^0\pi^+\pi^-$ decays, for instance, $\pt \lesssim 133$ MeV. In reality, however, the transverse momentum of the $\pi^0$ is not directly measured, but rather inferred from the energy and position of the $\pi^0\to \gamma\gamma$ photons detected in the ECAL~\footnote{The photon direction is in fact measured as a preliminary step in the analysis, although with a poor $Z$ vertex resolution. This is done in order infer the actual position of incidence of the photon on the surface of the ECAL and to the best of our knowledge is not used in the later stages of the analysis. We assumed this has negligible impact in our discussion.}. This is done by assuming the two photons to come from a pion, and computing the opening angle $\theta$ between their momenta as
\begin{equation}\label{eq:angleformula}
   \cos{\theta} = 1 - \frac{m_{\pi^0}^2}{2E_{\gamma_1}E_{\gamma_2}}.
\end{equation}
Once $\theta$ is known, the $Z_{\rm vtx}$ position of the $\pi^0$ decay, which coincides with that of the $K_L$, is calculated by assuming that the decay occurred precisely at a transverse position of $(X=0,Y=0)$. Note that for a fixed distance between the photons in the ECAL, a smaller $\theta$ implies a smaller $Z_{\rm vtx}$ value (further from the ECAL). Two key assumptions in this strategy are that the photons come from a $\pi^0$ and that the decay occurred exactly along the center of the beam. The prediction for $K_L$ transverse momentum is small, and a direct measurement is performed with $K_L\to 3\pi^0$ and $K_L\to 2\pi^0$, while the measurement of $K_L\to \gamma \gamma$ reconstructs $\ptK$ by, again, assuming the decay to happen at $(X=0,Y=0)$. The beam size can also be directly measured in the multi-pion final states, where it is found that the distributions in $Y$ are well approximated by a box function with $8$~cm width. We show both the $\ptK$ and the beam size distributions in \refapp{app:pt}, as well as a simple analytical fit to those which we use in our simulations for convenience.

\subsection{Reconstruction of new physics signatures}

\begin{figure}[t]
    \centering
    \includegraphics[width=0.49\textwidth]{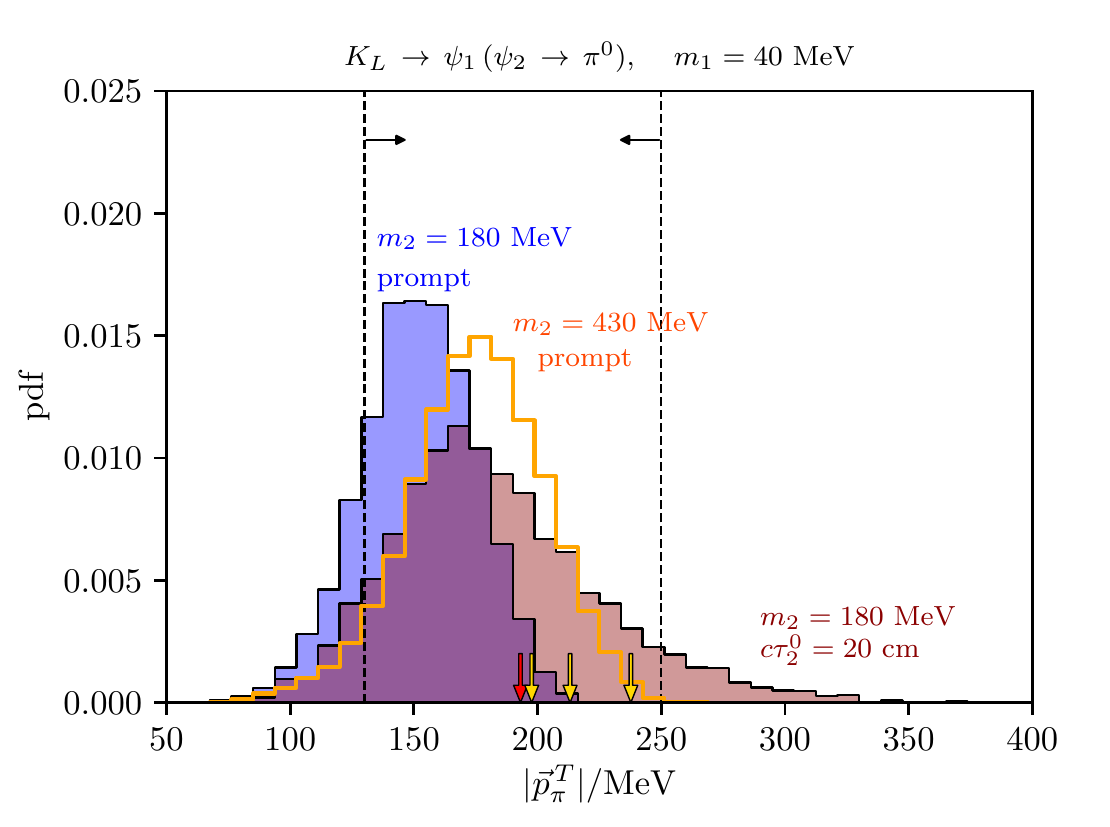}
    \includegraphics[width=0.49\textwidth]{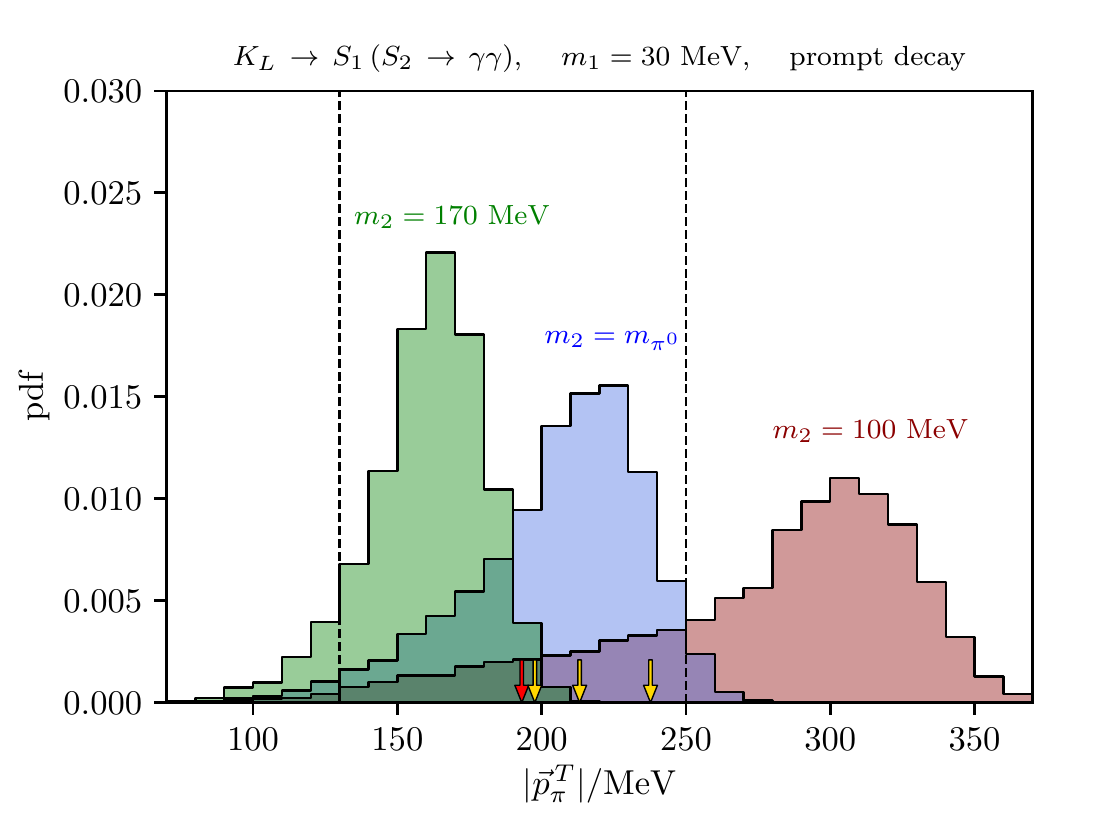}
    \includegraphics[width=0.49\textwidth]{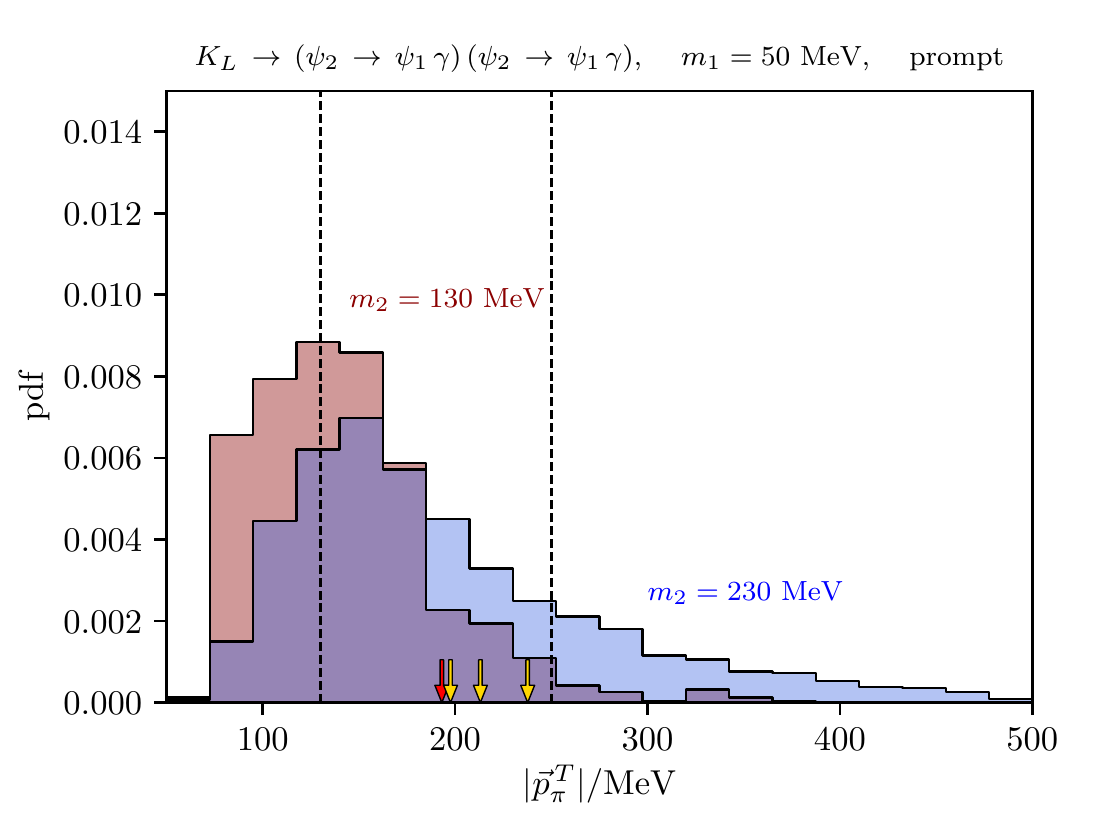}
    \caption{The distribution in $\pt$ for four different scenarios. \textbf{Top right:} scenario (A) with a prompt or long-lived $\psi_2$ Majorana fermion. \textbf{Top left:} scenario (B) for three different choices of a $\pi^0$ impostor mass. \textbf{Bottom:} scenario (C) with different $m_2$ masses.}
    \label{fig:kinematics}
\end{figure}

The kinematics of our signal can be significantly different from the $K_L\to\pi^0\nu\overline{\nu}$ three-body decays, and we would like to understand its impact on the KOTO signal selection. Before discussing reconstruction, however, let us first remark on a direct prediction that can be obtained in case (A). Since we produce two massive states with potentially hundreds of MeV in mass, the typical transverse momentum of $X_2$, and therefore of the emitted $\pi^0$, cannot be too large. Ignoring beam size and $\ptK$, the true pion transverse momentum is bounded by
\begin{align}
\pt_{\rm true} &<  \frac{m_{K_L}}{4}\left[(1+r_2^2-r_1^2)\lambda^{1/2}(1,y_1^2,y_\pi^2) + (1+y_\pi^2-y_1^2)\lambda^{1/2}(1,r_1^2,r_2^2) \right],
\end{align}
where the maximum is independent of $m_2$ and is attained when setting $m_1$ to be vanishingly small,
\begin{align}
\pt^{\rm max}_{\rm true} = \frac{m_{K_L}}{2}\left(1-\frac{m_\pi^2}{m_{K_L}^2}\right) \simeq 230\, \text{MeV with } m_1=0.
\end{align}
It is clear that before experimental smearing, the masses of $X_1$ and $X_2$ contain a direct prediction for maximum $\pt$. Now let us investigate how this changes in the reconstruction procedure.

Two novel possibilities for mis-reconstruction arise in our pair production models. Firstly, the $X_2$ particle may have a finite lifetime inside the detector and may decay at a transverse distance from the beam such that $\sqrt{X^2+Y^2}>$ (beam transverse size). In principle, all cases considered in this paper allow for macroscopical lifetimes of $X_2$, although constraints from beam dump experiments must be taken into account for any specific model. Secondly, the invariant mass of the diphoton signature may be significantly different from that of a $\pi^0$, being either peaked at $m_{\gamma\gamma}=m_{S_2}$ in scenario (C), or broadly distributed as in scenario (B). Both of these possibilities are already well-known and are intrinsic to certain backgrounds, such as neutron produced $\eta\to \gamma\gamma$ decays (CV-$\eta$), and $K_L\to\gamma\gamma$ decays from scattered-$K_L$ mesons~\cite{Su:2019xfn,Su:2017dlx,Su:2018vql}. 

To study the impact of our fake $K_L\to\pi^0\overline{\nu}\nu$ signals on the reconstruction, we developed a toy Monte Carlo (MC) simulation of the experiment (details in \refapp{app:pt}). We implement Gaussian detector resolutions and selection criteria as detailed in the analysis of 2015 data~\cite{Ahn:2018mvc} with the modifications in $\pt$ and $Z_{\rm vtx}$ cuts as introduced in the 2016-2018 unblinding talk~\cite{kotoKAON19}. \reffig{fig:kinematics} shows our MC predictions for scenarios (A) to (C) after all analysis cuts have been performed (except for $\pt$ cuts). We enclose the signal region with dashes, and show the observed events with arrows. Yellow arrows stand for events that are not consistent with backgrounds at this moment, and the red arrow represents the event that is compatible with (reevaluated) backgrounds. Clearly, a lot more possibilities in terms of $\pt$ distributions arise with pair production of dark states. Scenario (A) demonstrates the long tail exhibited by new states that decay within tens of cm inside the detector, even if the true pion momentum is very forward. A similar tail is observed in the dipole decay case, although mostly due to the fact that the gammas are uncorrelated with each other. In scenario (C), masses close to and lighter than the pion masses are preferred if $S_2$ is short-lived. Otherwise, the distribution develops a similar tail to that observed in case (A).
%

\subsection{\(Z\!-\!Z^\prime\) mixing parameter space}

\begin{figure*}[t]
    \centering
    \includegraphics[width=0.49\textwidth]{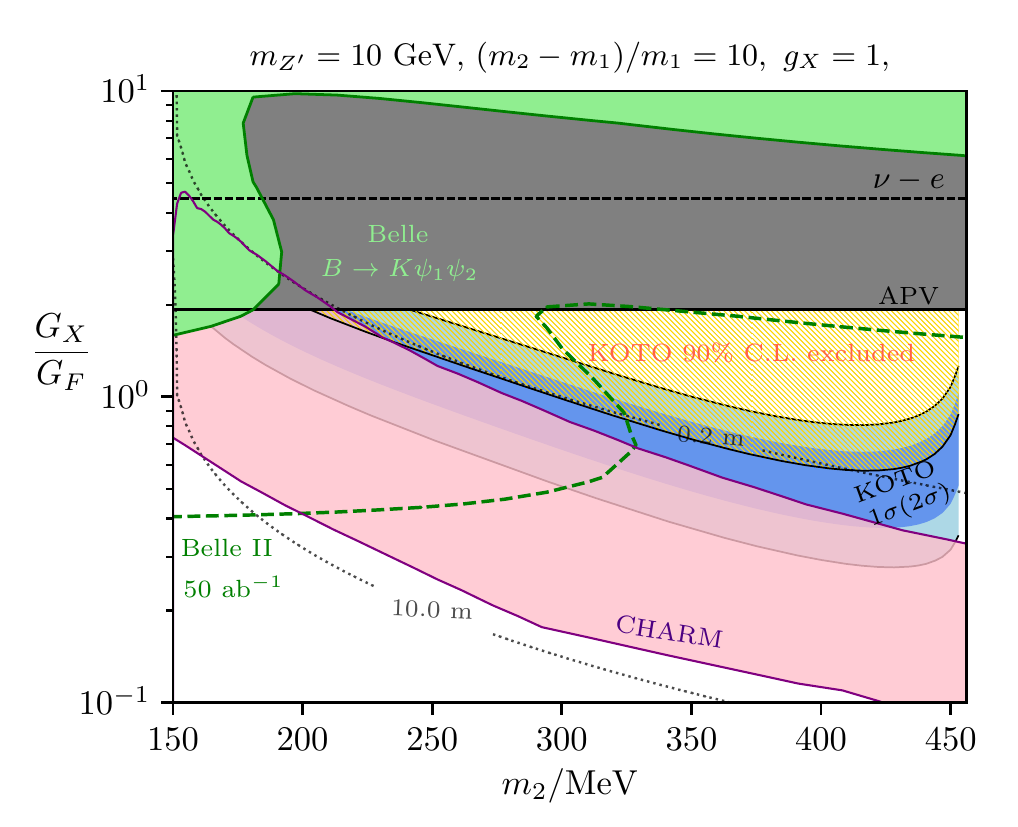}
    \includegraphics[width=0.49\textwidth]{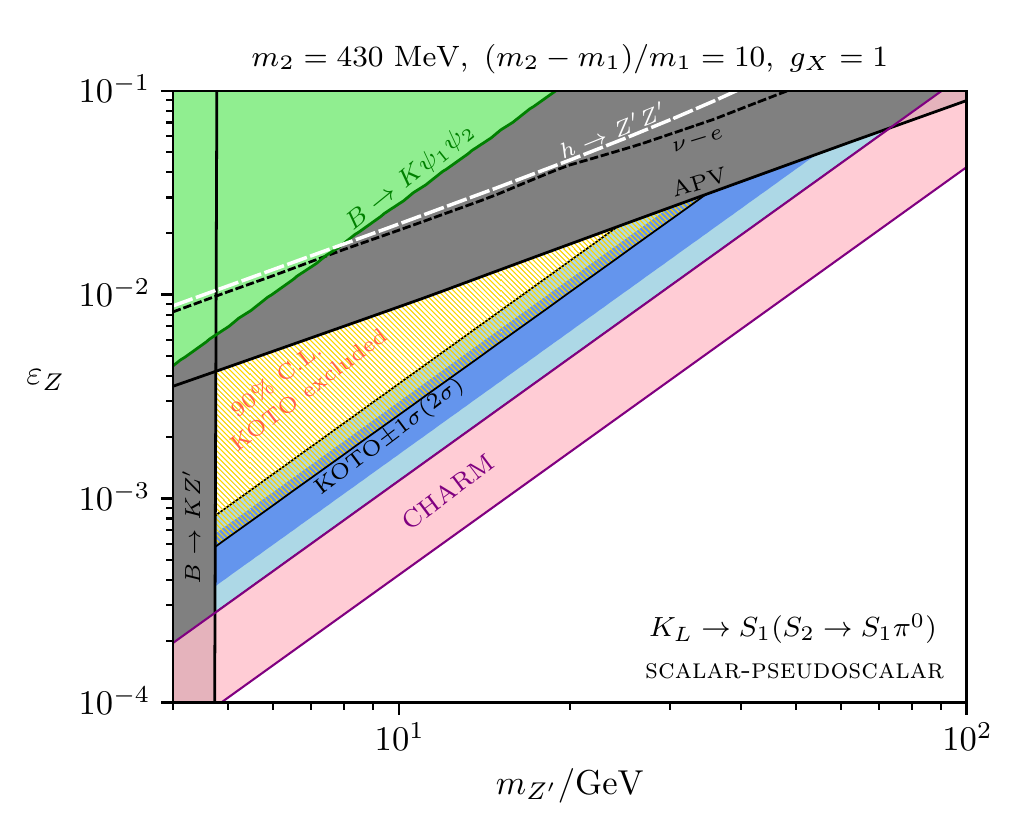}
    \caption{Parameter space for $Z\!-\!Z^\prime$ mixing and dark scalars. \textbf{Left}: the preferred band where three KOTO events can be explained at 1$\sigma$ (dark blue) and $2\sigma$ (light blue) in the phenomenological coupling $G_X$ versus $m_1$ plane. The green region is excluded by $B\to K\overline{\nu}\nu$ searches, and the red dashed line corresponds to a total BR$(B\to K \psi_1 \psi_2) > 1.2 \times$BR$(B^+\to K^+\overline{\nu}\nu)_{\rm SM}$, twice as large as the Belle-II sensitivity. \textbf{Right}: Same as the left plot, but in the $\varepsilon_Z$ versus $m_{Z^\prime}$ plane. \label{fig:plotscalars}}
\end{figure*}

Now that we have explored the kinematical properties of our signal, we turn to investigating the available parameter space. To provide a concrete example, we do this only for the case of a vector portal via $Z\!-\!Z^\prime$ mixing with off-diagonal couplings to dark fermions and scalars, as production and decay of dark states is fixed by the same coupling combination in this case. The relevant Lagrangians are \refeq{eq:XEFTmassmixing} combined with \refeq{eq:Lfermionpair} for dark fermions, and \refeq{eq:Lscalarpair} for dark scalars. We plot the regions of preference to explain 3 events at KOTO as well as upper limits on the $K_L$ BR in our new physics model according to
\begin{equation}
    \text{BR}(K_L \to \pi^0 \overline{\nu}\nu)_{\rm KOTO} =  \text{BR}(K_L \to X_1 X_2)_{\rm KOTO} \times \frac{\epsilon_{\rm NP}}{\epsilon_{\rm SM}} \times \left( 1- e^{-\langle L \rangle /\langle\ell_{\rm dec}\rangle} \right) \times \text{BR} (X_2 \to X_1 \pi^0)
\end{equation}
where $ \text{BR}(K_L \to \pi^0 \overline{\nu}\nu)_{\rm KOTO}$ is either the preferred BR in order to explain the anomalous events, or the upper limit quoted by KOTO. Here, $\epsilon_{\rm NP}$ is the signal selection efficiency in the new physics model (not taking into account the probability of $X_2$ to decay outside the detector). The ratio between the new physics and the SM signal selection efficiencies within our simulation, $\epsilon_{\rm NP}/\epsilon_{\rm SM}$, is shown in \refapp{app:pt}, and is implemented via a third degree polynomial fit in our plots. We also define $\langle \ell_{\rm dec} \rangle$ as the average decay length of the new states in the KOTO experiment, and $\langle L \rangle$ as the average decay distance available for $X_2$ inside the KOTO experiment. For simplicity, we take $\langle L \rangle = 1.5$~m and $\langle E_{X_2} \rangle = 1.0$~GeV at KOTO. Note that this procedure neglects finite lifetime effectcs on the signal reconstruction, which is a good approximation for the regions of parameter space where $X_2$ is short-lived ($c\tau_2^0 < 1$~cm).

Fig~\ref{fig:plotscalars} shows the parameter space for a pair of scalars. The region in blue shows where the model is compatible with the observation of 3 events at KOTO at $1\sigma$ and $2 \sigma$. In this case, $K_L$ decays prefer large $G_X$ values and $X_2$ is typically short-lived. We find that lowering $m_2$ to below $200$~MeV implies much larger couplings for explaining KOTO excess, and is mostly excluded in this $Z^\prime$ model due to beam dump constraints. $B$ meson decay constraints are also shown, but are weak in comparison with direct constraints on $Z\!-\!Z^\prime$ mixing.

\begin{figure*}[t]
    \centering
    \includegraphics[width=0.49\textwidth]{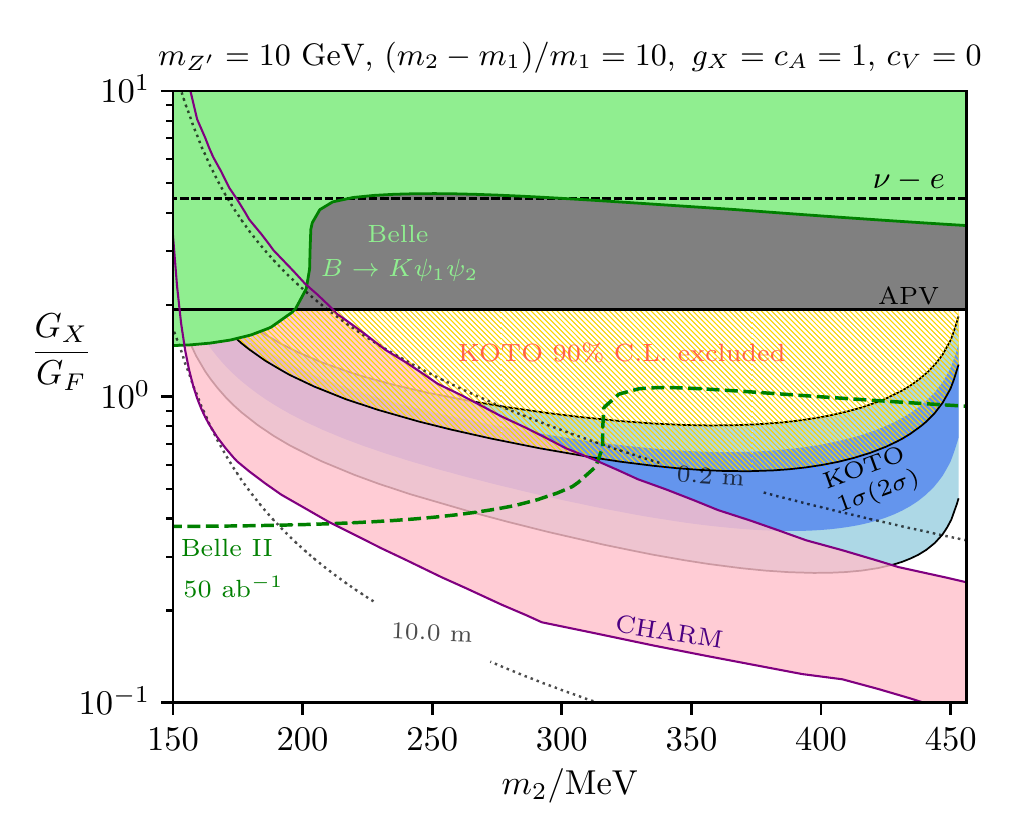}
    \includegraphics[width=0.49\textwidth]{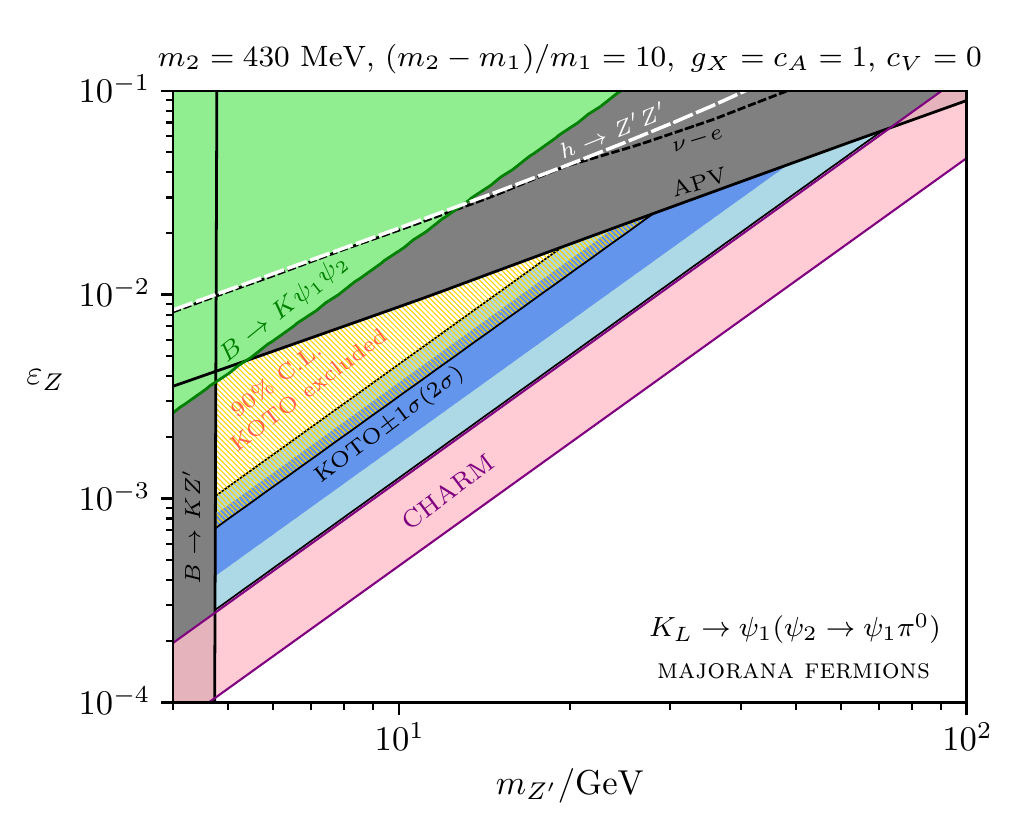}
    \caption{Parameter space for $Z\!-\!Z^\prime$ and dark Majorana fermions. \textbf{Left}: the preferred band where three KOTO events can be explained at 1$\sigma$ (dark blue) and $2\sigma$ (light blue) in the phenomenological coupling $G_X$ versus $m_1$ plane. The green region is excluded by $B\to K\overline{\nu}\nu$ searches at $90\%$~C.L., and the red dashed line corresponds to the Belle-II sensitivity to this channel at $90\%$~C.L. \textbf{Right}: Same as the left plot, but in the $\varepsilon_Z$ versus $m_{Z^\prime}$ plane.For Dirac fermions, the bounds on $\varepsilon_Z$ do not change, but all meson decay and lifetime curves are multiplied by $\sqrt{2}$.
    \label{fig:plotfermions}}
\end{figure*}

We show our results for the case of $K_L$ decay to fermions in \reffig{fig:plotfermions}. As we have seen before, the enhancement in $K_L\to\psi_1\psi_2$ scales slightly differently in $m_2$ than in the scalar case, and so larger $m_2$ values are less tightly constrained by KOTO and beam dumps. In addition, $B$-decay constraints become more relevant, but not yet particularly sensitive to the allowed region. For lighter $m_2$, the bounds become stronger, but we find that as long as $c\tau_{\psi_2}^0 < 20$ cm, it is in an interesting region for KOTO, be it due to new constraints or explaining the observed events. 

We note that regardless of recent excess events, the results of KOTO $K_L\to \pi^0\slashed{E}$ provide strong restrictions on the parameter space of these models (yellow regions in Figures \ref{fig:plotscalars} and \ref{fig:plotfermions}).
This sensitivity can be directly attributed to the two-body nature of the 
$K_L\to X_1X_2$ decays, and correspondingly large rates. The edge of the excluded region can indeed become an interesting frontier of the dark sector physics in the future, as more experimental understanding is gained into the origin of the excess. 

We now discuss other sources of constraints in the parameter space of the $Z-Z^\prime$ mixing models we consider. At large $Z^\prime$ masses, the strongest experimental constraints on $\varepsilon_Z$ come from atomic parity violation (APV)~\cite{Porsev:2009pr,Bouchiat:2004sp,Dror:2018wfl}, neutrino-electron scattering ($\nu-e$)~\cite{Bilmis:2015lja}, and higgs decays~\cite{Sirunyan:2018owy,Aaboud:2019rtt}. We take these constraints from Ref.~\cite{Dror:2018wfl}. At low energies, the bounds become more severe due to searches for $B\to K Z^\prime$~\cite{Grygier:2017tzo,Aubert:2008ps,Aaij:2016qsm} and $K\to\pi Z^\prime$~\cite{Artamonov:2008qb,AlaviHarati:2003mr} decays, so we restrict our discussion to rather heavy mediators, namely $m_{Z^\prime}$ in the mass range $\sim 10$ GeV. Additional decay channels for $X_2$ and the three-body $B$ decays are discussed in \refapp{app:threebody}.

\bm{$B$}\textbf{ factories}: We emphasize that two body decays $B\to X_1 X_2$ are much suppressed due to the small decay constant, $f_B\ll m_B$, which is not the case in $B\to {\rm M} X_1 X_2$, with ${\rm M}=\pi,K,K^*$. Despite being a three-body process, both neutral and charged $B$ mesons contribute and the rates can be large since $|V_{tb}^*V_{ts}|\gg|V_{ts}^*V_{td}|$ (see \refapp{app:threebody}). This counts as signal in $B\to K \overline{\nu}\nu$ searches if $X_2$ appears invisible by either decaying invisibly (out of the detector acceptance or when $X_2$ has a finite BR into invisible, such as $X_2\to X_1 \overline{\nu}\nu$), or by escaping the detector volume. B factories operate at higher energies than KOTO, so $X_2$ has a larger chance of escaping the detector. If $X_2$ decays inside the detector most of the time, one may design a dedicated search to look for the decays that do happen inside the volume, although such $\pi^0$-like signatures with missing energy are usually challenging and will not be discussed here. In drawing our curves we do not attempt a complete simulation of B factories experiments, but compute current constraints and future sensitivity by assuming a typical detector length of $\langle L_{\rm det} \rangle = 2$~m at Belle and Belle-II. Note that the constraint $m_1+m_2<m_K$ implies that the masses of the dark states have little impact on the missing invariant mass $m_{\nu\nu}$ in $B$ decays. The projected sensitivity of Belle-II to $B\to K \overline{\nu}\nu$ decays is 10\% of the SM value~\cite{Kou:2018nap}.

\textbf{Beam dumps}: Beam dump constraints can arise via production of dark states via the off-shell mediator. As it proceeds via an effective dimension-6 operator with weak-scale strength, then both bremsstrahlung $f^+f^-\to f^+f^- \psi_1 \psi_2$ and direct $qq \to \psi_1 \psi_2$ production are not very effective. Instead, $\eta$ decays of the type $\eta \to \psi_1 \psi_2$ dominate. The most sensitive experiments in the regime of interest are CHARM and NuCal. CHARM has published limits on light scalars produced at the target and decaying into $\gamma\gamma$, $e^+e^-$, and $\mu^+\mu^-$~\cite{Bergsma:1985qz}, and later works have adapted searches for heavy neutral lepton decays into $\nu e^+e^-$ final states to generic dark sectors~\cite{Gninenko:2012eq,Ilten:2018crw,Tsai:2019mtm,Darme:2020ral}. \refref{Darme:2020ral} has surveyed constraints on dimension-6 ``fermion portal" operators of the type $(g/\Lambda^2) \overline{\psi_2}\gamma_\mu\gamma^5\psi_1 \overline{f}\gamma^\mu\gamma^5 f$, and their results for a Z-aligned scenario can be adapted to our models with a few caveats. Firstly, the beam dump constraints were obtained by re-scaling the limits in Ref.~\cite{Tsai:2019mtm}, which assumes small mass-splitting between $\psi_1$ and $\psi_2$ and final state leptons only. Secondly, no coupling to neutrinos was assumed in their case, but as shown in \refapp{app:threebody} the invisible BR for $\psi_2$ is of the order of 10\%, and 5 times larger than that into $\psi_1 e^+e^-$. Finally, we show results for Majorana fermions, so factors of two for production and total $\psi_2$ lifetime have to be taken into account. 
Typical beam dump bounds represent a band in the parameter space. 
The ``low coupling" boundary is quite sensitive to details about efficiencies and $\eta$ meson distributions, but for us the ``large coupling" boundary is the most relevant. This one is predominantly sensitive to the total lifetime of $X_2$, and in that sense it is robust. NuCal is expected to set somewhat stringer lower limits on our parameter space than CHARM~\cite{Blumlein:1990ay}, but by a factor smaller than two.
Future experiments such as SHiP~\cite{Anelli:2015pba,Alekhin:2015byh} and DUNE~\cite{Acciarri:2016crz,Berryman:2019dme} can provide an intense source of $\eta$ mesons where the subsequent decay of $X_2$ may be searched for. Moreover, the models presented here can also be probed via collider production of $\eta$ mesons with largest attainable boosts. In particular, the planned experiment FASER~\cite{Ariga:2018zuc} may present a good avenue for searches of the displaced $X_2\to X_1\pi^0$ decays. 

\textbf{Higgs decays}: At the LHC, the Higgs boson production and decays to $Z^\prime$ particles is a sensitive probe of our vector portal due to longitudinal emission of the new vector. In our dark sector, however, $Z^\prime$ decays mostly to $X_1 X_2$ particles, and therefore the usual $h\to (Z^\prime\to\ell^+\ell^-) (Z^\prime\to\ell^+\ell^-)$ and $h\to Z (Z^\prime\to\ell^+\ell^-)$ searches are relaxed by the small BR of $Z^\prime\to\ell^+\ell^-$. Instead, we place bounds using the constraint $\Gamma_{h\to Z^\prime Z^{(\prime)}} < 0.1 \times \Gamma_h^{\rm SM}$, avoiding large modifications to the Higgs BRs. This is a model-dependent constraint, as explained in \refref{Dror:2018wfl}, but does not lead to any significant restriction to the parameter space of interest. In \reffigs{fig:plotfermions}{fig:plotscalars}, we fix  $m_{\rm UV}=500$ GeV. Note that $h\to Z^\prime Z$ is much weaker than $h\to Z^\prime Z^\prime$, despite scaling as a lower power in $\epsilon_Z$.

\bm{ $Z^\prime\!-\!X_1\!-\!X_1$} \textbf{vertex}: We have only discussed the off-diagonal coupling between $Z^\prime\!-\!X_1\!-\!X_2$ so far. In generic UV completions, we expect diagonal couplings would also arise, but these can be easily suppressed. A Majorana mass term from the condensation of a dark Higgs, for instance, can be invoked to split the components of a vector-like fermion. In the mass basis, $\mathcal{L} \supset g_X (c_{ij}/2) \, Z^\prime_\mu \overline{\psi_i}\gamma^\mu \gamma^5 \psi_j + {\rm h.c.}$, and one can achieve $c_{11} \ll c_{12}$ by assuming maximal mixing, in analogy with inelastic dark matter models~\cite{TuckerSmith:2001hy}. If $\psi_1$ is light, one may worry about $\pi^0\to \overline{\psi_1} \psi_1$ decays, as these are strongly constrained by NA62~\cite{na62KAON19}, BR$(\pi^0\to$inv$)<4.4\times10^{-9}$ at 90\% C.L. In this Majorana fermion model, however, we find
\begin{equation}
    \text{BR}(\pi^0\to \overline{\psi_1} \psi_1 )_{\rm M} = 3.7 \times 10^{-9} \times c_{11}^2 \left(\frac{G_X}{G_F}\right)^2 \left(\frac{r}{0.25}\right)^2 \lambda^{1/2}\left(1,r^2,0\right).
\end{equation}
with $r=m_1/m_\pi$. Clearly, for all $r<0.5$ this suggests a rather weak constraint on $c_{11}$ for the parameter space of interest. Note that, in a model with sizeable $c_{11}$, the decay $\psi_2\to \psi_1 \psi_1 \overline{\psi_1}$ needs to be taken into account, and will dominate if kinematically accessible and if $\epsilon_Z^2 g^2/4c_W \ll c_{11}^2 g_X^2$. If comparable with the BR into visible, such decays actually relax the beam dump constraints on our scenario, and one is free to explore previously excluded values of $m_2$. We illustrate this point in the next section within scenario (A), both for the current model and for the case of long-distance $\Delta S = 1$ pair production.

\subsection{Invisible decays in $Z\!-\!Z^\prime$ mixing and long-distance $\Delta S =1$ scenarios}

\begin{figure}
    \centering
    \includegraphics[width=0.49\textwidth]{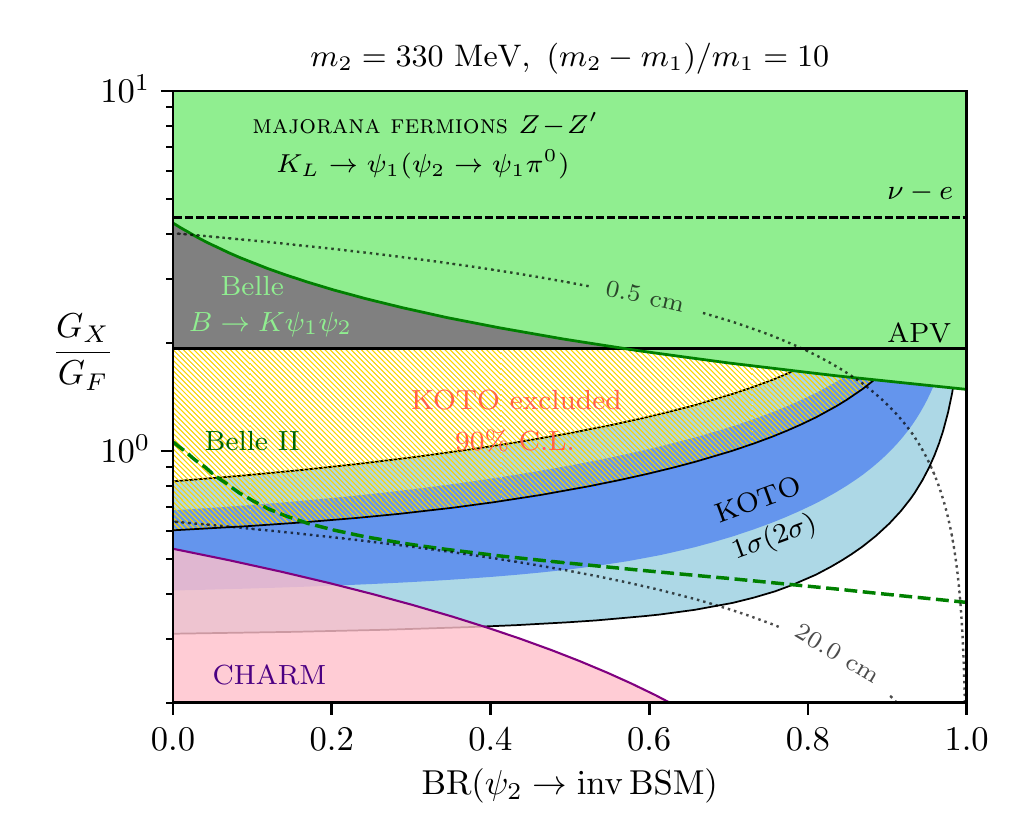}
    \includegraphics[width=0.49\textwidth]{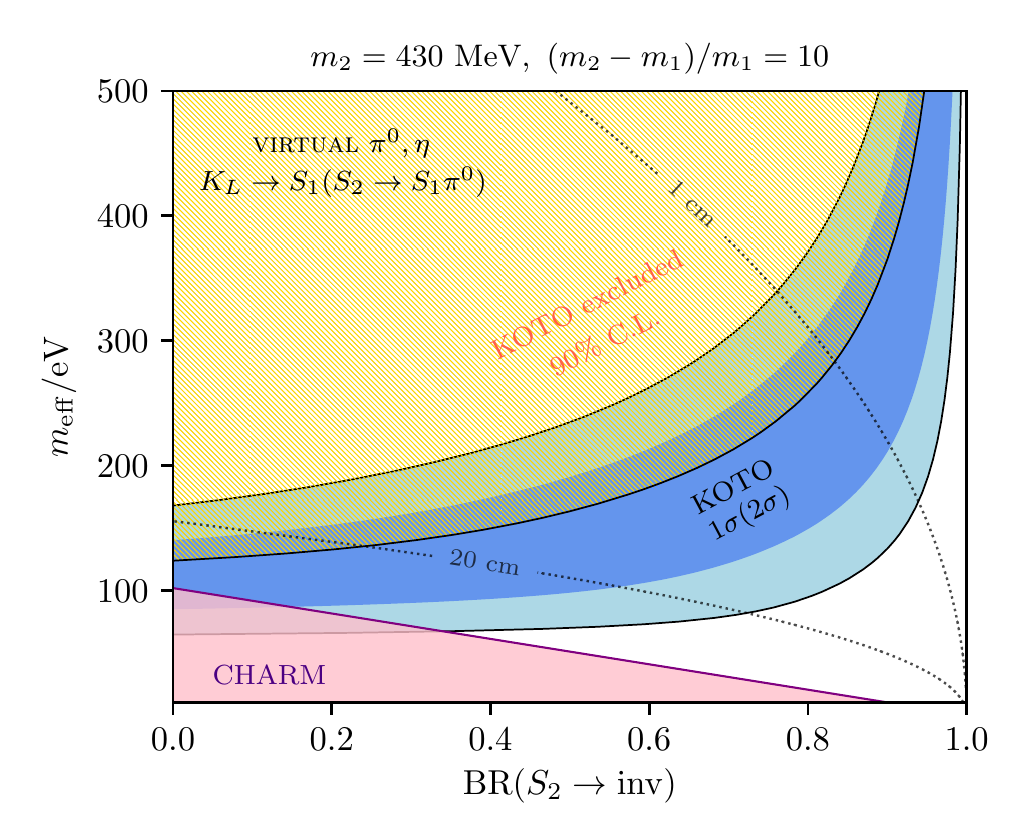}
    \caption{The change in parameter space when $X_2$ gains a new invisible decay channel with respect to the benchmark model. \textbf{Left:} The majorana fermion case in the $Z\!-\!Z^\prime$ mixing model. \textbf{Right:} The virtual $\pi^0$, $\eta$ model with dark scalars. 
    }
    \label{fig:virtualpi0}
\end{figure}

Our generic dark sectors may predict a sizeable BR($X_2 \to $ inv), where new invisible dark states can be produced in the final state. This impacts the relevant parameter space for $K_L$ decays and relaxes beam dump constraints, as the $X_2$ lifetime shortens. We explore this possibility in \reffig{fig:virtualpi0}, first in the case of a $Z\!-\!Z^\prime$ mixing model and Majorana fermions, where we allow for an arbitrary BR into invisible dark states. This can be easily achieved by controlling the decay rate for $\psi_2 \to \psi_1 \psi_1 \psi_1$, for instance. In defining BR$(\psi_2\to$ inv BSM), we do not include the rate for $\psi_2\to \psi_1 \overline{\nu}\nu$. Unsurprisingly, even for moderate values of invisible BR the beam dumps no longer constrain the enhancement in $K_L$ decays, and the preference region for KOTO remains somewhat unchanged. As a consequence, $B \to K \slashed{E}$ searches also become more sensitive, as $\psi_2$ appears as missing energy in the detector more often. In this sense, the Belle-II coverage of the $K_L$ enhancement in our models is much broader than what \reffigs{fig:plotscalars}{fig:plotfermions} may suggest.

Now, consider the case where new particles are produced in $K_L$ decays through a combination of flavor-diagonal couplings and long-distance $\Delta S = 1$ operators. As discussed in \refsec{sec:virtualpi0}, this case is very effectively constrained by beam dumps due to the finite lifetime of $S_2$. However, if $S_2$ can decay invisibly, its lifetime will shorten and the beam dump constraints are relaxed. This effect is shown on the right panel of \reffig{fig:virtualpi0}, where we pick a $430$~MeV $S_2$ particle with a visible BR into $\pi^0$ as fixed by \refeq{eq:S2pi0decay}. We then assign it an arbitrary BR into invisible, BR$(\to{\rm inv})$, and ignore any other decay channel for $S_2$ for simplicity. As above, the CHARM constraints get significantly weakened, even for moderate BR$_{\rm inv}$, as the lower limit on $m_{\rm eff}$ is exponentially sensitive to the total lifetime. For comparable BR$_{S\to \pi^0 S_1}$ and BR$_{\rm inv}$, we see only a marginal change in the number of events at KOTO, as $S_2$ has a smaller chance to escape the detector but an increasingly larger change to decay invisibly. The dependence of our argument on $m_2$ is also marginal, where larger $m_2$ values are somewhat less constrained by beam dumps.

To conclude this section, we emphasize that due to the visible decay nature of our pair production hypothesis, there is no lower bound on the lifetime of $X_2$. In fact, shortening the lifetime of $X_2$ can be easily accomplished by allowing for invisible decays. This is not the only option, of course, as scenarios (B) and (C) already require independent couplings for production and $X_2$ decay, and avoiding beam dump constraints in those cases is an even easier task.

\subsection{Comparison with recent literature}

Proposals to enhance $K_L$ decays with $\overline{\nu}\nu$ pairs have focused purely on production of \emph{invisible} particles in either 2-body or 3-body decays. We use our own simulation to understand the differences in the $\pt$ distributions with respect to our visibly decaying dark state scenarios.
For instance, consider the proposals in Refs.~\cite{Kitahara:2019lws,Egana-Ugrinovic:2019wzj}, where a new light scalar $S$ mixes with the Higgs, being produced via $K_L\to \pi^0 S$. As pointed out in Ref.~\cite{Fuyuto:2014cya,Hou:2016den}, if $m_S\simeq m_\pi$ or $m_S>2m_\pi$, then such scenarios can evade bound \cite{Grossman:1997sk} since $K^+\to\pi^+ S$ is unobservable on top of large $K^+\to\pi^+\pi^0$ or $K^+\to\pi^0\pi^+\pi^-$ backgrounds, respectively, since $S$ comprises all of the missing energy. In fact, ignoring beam size and $\ptK$, the maximum transverse momentum of the pion in those cases is simply 
\begin{align}
   \pt_{\rm true} < \frac{1}{2 m_{K_L}} \lambda^{1/2}(m_{K_L}^2,m_\pi^2,m_S^2) \simeq   \left\{
\begin{array}{ll}
       209\, \text{MeV for } m_S = m_\pi,
       \\
       133\, \text{MeV for } m_S = 2m_\pi.
\end{array} 
\right.
\end{align}
Even after reconstruction, it is clear that the second option leads to transverse momenta that are too small, below the experimental cuts to remove $K_L\to\pi^0\pi\pi$ backgrounds. To understand to what degree this upper bound is valid, we simulate such decays and show that for the $m_S\simeq m_\pi$ option, once reconstruction effects are included one can, in fact, explain all observed events if $S$ has a mass not much greater than $m_S\gtrsim150$ MeV. This is to be compared with our scenario (C), which is essentially a generalization of the $K_L\to\pi^0 S$ signature. In that case, the mass of the invisible state is a free parameter, and one can achieve much larger phase space for the $K_L$ decays. 
\begin{figure}[t]
    \centering
\includegraphics[width=0.49\textwidth]{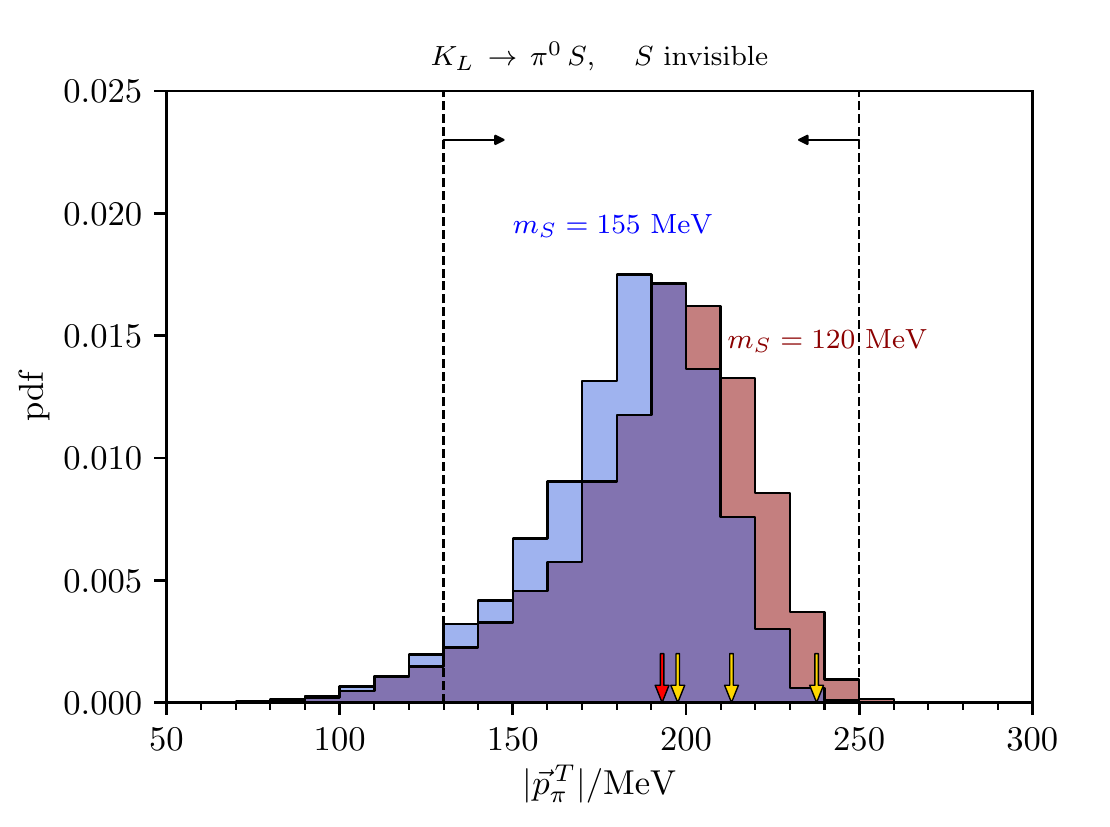}
\includegraphics[width=0.49\textwidth]{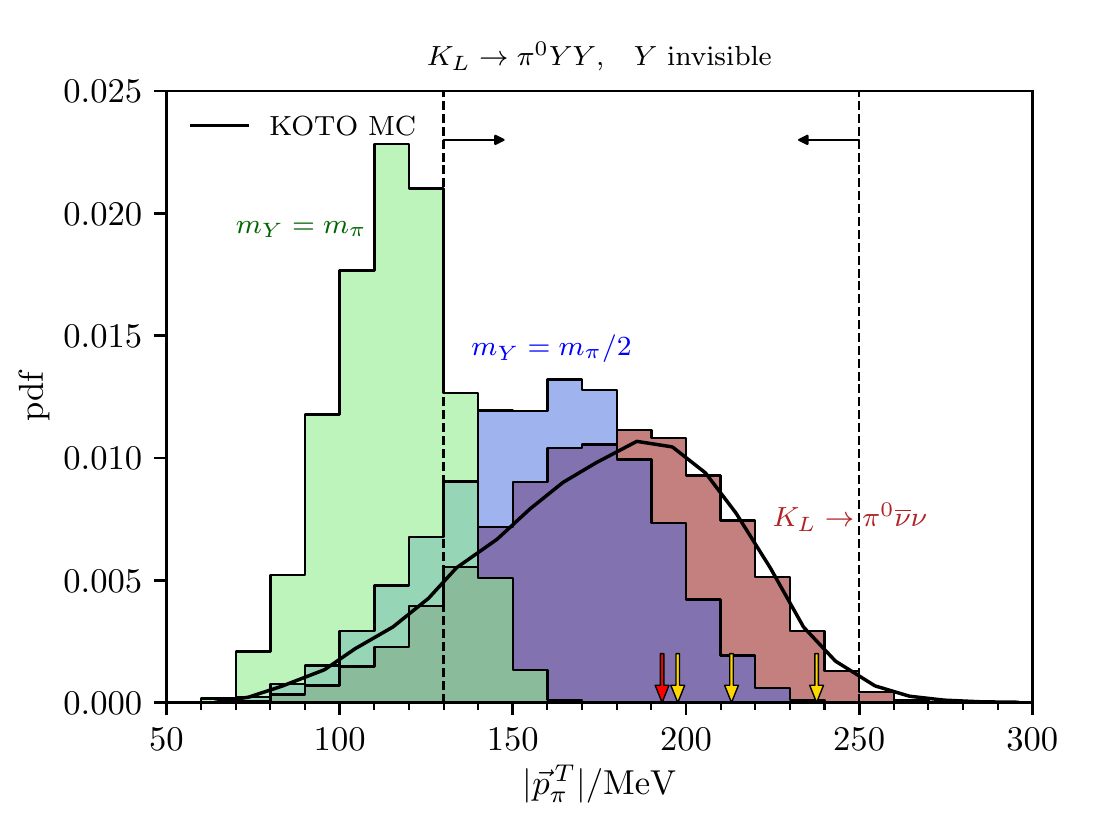}
    \caption{Area normalised $\pt$ distribution in the SM decay and other BSM scenarios. \textbf{Left}: $K_L \to \pi^0 S$ events at KOTO for the case where $m_S$ is close to $m_\pi$. \textbf{Right}: $K_L\to\pi^0 Y Y$ for the SM signal compared with BSM scenarios where a pair of invisible $Y$ particles are produced. The solid black line corresponds to the KOTO MC from Ref.~\cite{Nakagiri:2019yec}, and helps validate our own simulation. A final $Z_{\rm vtx}$ cut has little impact on the distributions, and is left out only in the SM histogram for a fair comparison with the KOTO MC.}
    \label{fig:comparisons}
\end{figure}

Another set of proposals involve the production of more than one invisible particle, $K_L\to\pi^0 YY$. In that case, similar considerations to the above can be made, where now one may apply formula \refeq{eq:3bodypt}. Immediately one can discard scenarios where $m_{K^+}-m_{\pi^+}<2m_Y<m_{K_L}-m_{\pi^0}$, as the pion would have virtually no phase space~\cite{Fabbrichesi:2019bmo}. Scenarios with $m_Y = m_\pi$ are also disfavored, as the maximum $\pt$ is precisely around the minimum $\pt$ in the signal region. Ref.~\cite{Ziegler:2020ize} has recently raised exceptions to the Grossman-Nir bound via a direct flavor violating coupling to new scalars, where either one or a pair of invisible particles is produced, with the lightest of which being a dark matter candidate. In that case, the bounds on $\pi\to YY$ with $Y$ invisible can be very stringent~\cite{na62KAON19} requiring that $m_Y/2>m_{\pi}$. In analogy with the 2-body decays, we can already deduce that if such scenarios are to reproduce the large $\pt$ events observed at KOTO, then $m_Y$ better not be too far from the $m_Y/2>m_{\pi}$ lower bound. We verify this in our simulation, and show the resulting distribution in \reffig{fig:comparisons}. This proves that such scenarios are indeed predicting large number of events in the signal region, provided $m_Y < m_\pi$. Note that if the observed events turn out to be confirmed, an explanation of the largest $\pt$ is very challenging unless $m_Y \lesssim m_\pi/2$.

\section{Connecting meson decays to light dark matter}\label{sec:UVcompletions}

In this section we would like to discuss the possibility of $X_1$ being a dark matter candidate within several UV complete model. 
We would like to open this section with a few observations of general nature. 

\begin{itemize} 

\item In all but one model considered in the previous sections $X_1$ can be stable 
and therefore provide a candidate for dark matter. It is easy to see that the exception is the model with a pion impostor, $X_2\to \gamma\gamma$, where the number of $X$ particles is not conserved, and $X_1$ will decay to light SM states. 

\item In all models with stable $X_1$, which we now promote to a dark matter candidate, the couplings of $X$ particles to the SM are large enough to ensure full thermal equilibrium as inevitable initial condition. 
Therefore, in order to be a successful dark matter candidate, 
$X_1$ would have to be able to lower its own abundance by annihilation, with the
rate at the freeze-out on the order of $\langle \sigma {\rm v} \rangle_{f.o.} \propto\, $1\,pbn$\times c$. Smaller annihilation rates would lead to the overproduction 
of dark matter. 

\item Consistency of main cosmological probes (BBN and the CMB) 
imposes additional requirements. The dark matter mass $m_{1}$ would have to be in excess of a few MeV \cite{Nollett:2013pwa}, and the annihilation rate at late times
would have to be significantly smaller than the freeze-out rate, 
$\langle \sigma {\rm v}\rangle_{\rm CMB}\ll \langle \sigma {\rm v} \rangle_{f.o.}$.

\item Crucial for our set of models, $X_1-X_2-\pi^0$ coupling is too small 
to drive the freeze-out (co-)annihilation, and therefore new model components would have to get introduced to ensure overall viability of these dark matter scenarios. 

\end{itemize}

Since the theory of light WIMPs is well understood, we will use a number of model-building solutions that were developed over the years. In particular, we would like to consider the following constructions: 
\begin{enumerate}
\item  ``Cannibal" dark matter. 
Self-interaction in the scalar $S_1(=X_1)$ sector, {\em e.g.}  $S_1^3$ or $S^4_1$
terms in the Lagrangian can lead to $3\to2$ and $4\to 2$ 
depletion of the number density of $S_1$ 
\cite{Carlson:1992fn,Pappadopulo:2016pkp}. 
It is well appreciated this process 
must be accompanied by heat exchange between the dark sector and the SM, 
which have to come from new interactions, {\em e.g.} $S^2_1 \overline{d}d$
terms in the effective Lagrangian that can be induced by $(H^\dagger \Phi+\Phi^\dagger H)S_1^2$ terms added to Eq. (\ref{Phi0}).

\item Secluded annihilation (see {\em e.g.} \cite{Pospelov:2007mp}). 
For the fermionic dark matter $\psi_1(=X_1)$, one can introduce a new
scalar mediator $\phi$ that has parity-conserving couplings to $\psi$, {\em e.g.}  $\overline{\psi_1}\psi_1 \phi $. In that case, and for $m_\phi < m_1$, $\psi_1\overline{\psi_1}\to 
\phi \phi $ annihilation proceeds in $p$-wave, which is ``safe" against the CMB constraints \cite{Slatyer:2015jla}. Subsequent pre-BBN decay of $\phi$ must be ensured, perhaps at the price of introducing a linear Higgs portal coupling, $\phi H^\dagger H$ and/or 
$\phi \Phi^\dagger \Phi$, which will result in fast $\phi$ decays, and will be inconsequential for flavor physics if the mixing angle is in the range of 
$10^{-5}-10^{-4}$. 

\item ``Forbidden" dark matter. In this scenario, an unstable 
mediator slightly heavier than 
$X_1$ is introduced, 
let us call it $\phi$, so that the $X_1X_1 
\to \phi\phi$ occur at modest exponential penalty at the freeze-out energies, that becomes 
progressively more stringent as the temperature drops \cite{DAgnolo:2015ujb,Cline:2017tka}. As a result, even the $s$-wave annihilation is perfectly safe from the CMB constraints. In some cases, the SM particles themselves can serve as such a mediator. For example, 
effective $S_1^2(\pi^0)^2$ coupling can lead to efficient depletion of 
$S_1$, in which case its mass scale would have to be chosen closer to $m_\pi$ 
({\em i.e.} in the 100\,MeV, not 10\,MeV, range).

\end{enumerate}

Now we elaborate on some of these ideas,  
and provide an explicit model that 
gives a viable dark matter candidate. We do it for the scenarios with the fermionic $\psi_1$ 
and bosonic dark matter $S_1$, by introducing new annihilation channels into unstable mediators:
\begin{eqnarray}
\label{FermDM}
{\rm Fermionic~dark~matter}&:& ~\psi_1 + \overline{\psi_1} 
\to \phi+\phi\to~{\rm SM},~~m_\phi<m_1~{\rm or}~ m_\phi >m_1
\\
{\rm Bosonic~dark~matter}&:& ~S_1+S_1 \to \phi+\phi\to~{\rm SM},~~ m_\phi >m_1.
\label{BosDM}
\end{eqnarray}
To source this annihilation we use the most straightforward couplings, such as 
\begin{equation}
   {\cal L} \supset  \lambda_\phi S_1^2\phi^2;~~   y_\phi \phi \overline{\psi_1}\psi_1.
\end{equation}
The decay of $\phi$ mediator can be achieved via its mixing with the Higgs: 
\begin{equation}
    {\cal L}_\phi \supset A_\phi \phi H^\dagger H ~\to ~ \theta \times \phi \sum O_h,
\end{equation}
where $\theta $ is the mixing angle, and $O_h$ are the usual SM operators the Higgs field couples to, $O_h = (m_f/v)\overline{f}f,..$. The value of the mixing angle 
in the $\theta^2 \sim 10^{-9}-10^{-8}$ is currently not challenged by any of the existing experiments \cite{Beacham:2019nyx}, yet providing fast enough decays of $\phi$ on the pre-BBN time scales. 

Notice that (\ref{FermDM}) and (\ref{BosDM}) imply different annihilation regimes. 
Annihilation of two fermions into two scalar bosons occur in the $p$-wave for parity-conserving interactions, and therefore there is no requirement 
for $\phi$ to be heavier than $\psi_1$. For the dark matter 
built out of $S_1$ fields, the annihilation is in the $s$-wave, and must be 
``forbidden", otherwise it violates CMB constraints. 

The annihilation rate of the nonrelativistic 
fermion-antifermion pair is given by
\begin{equation}
    \sigma {\rm v}  = {\rm v}^2 \frac{3y_\phi^4}{64\pi m_1^2}\,F(m_\phi^2/m_1^2) \simeq 1\,{\rm pbn}\times c \times \frac{\rm v^2}{0.1}\times \left(\frac{y_\phi}{0.01}\right)^4 
    \left(\frac{\rm 100\,MeV}{m_1}\right)^2 
\end{equation}
Here $F(x) =(1-\frac{1}{2}x)^{-4}(1 -\frac89x+\frac29x^2)(1-x)^{1/2} $, and the last relation is taken at $m_\phi\ll m_1$. Parameter ${\rm v}$ here stands for the velocity of annihilating $\psi_1$, and is normalized to its typical freeze-out value. As is well-known, the annihilation rate of 1 pbn ensures correct dark matter abundance, and in this model this can be achieved with moderately small values of 
Yukawa couplings, $y_\phi \sim 10^{-2}$. 

The annihilation cross section in case of the forbidden 
bosonic dark matter calculated in the center-of-mass frame of colliding scalars of energy $E_1$, and has the value
\begin{equation}
\label{forbidden}
    \sigma{\rm v} = \frac{\lambda^2_\phi}{32\pi E_1^2}\times \sqrt{1-m_\phi^2/E_1^2}.
\end{equation}
The required size of $\lambda_\phi$ is exponentially sensitive to $\Delta m = m_\phi -m_1$. Indeed, the thermal average of rate (\ref{forbidden}) is suppressed by 
$\exp\{-2 \Delta m/T\}$, which for freeze-out temperature of $T\sim 0.05 m$ corresponds to $\exp\{-40\Delta m /m \}$. Requiring $\lambda_\phi<1$ results in a mild constraint on $\Delta m$, $\Delta m <0.6 m $. Thus we conclude that both 
(\ref{FermDM}) and (\ref{BosDM}) scenarios can be implemented rather broadly, without a fine-tuned choice of parameters, but at the expense of new ingredients in the model. 

These dark matter scenarios can be ``merged" with the discussion of meson decays, $K_L\to X_2X_{1(2)}$, and we would like to argue that mediator $\phi$ can be actually motivated from the point of view of a more fundamental theory. For example, in $Z'$-based models, we operate with dark currents such as $X_1\partial_\mu X_2 - X_2\partial_\mu X_1$, that are not conserved on account of $m_1\neq m_2$. These mass splittings cannot be fundamental, and are likely the result of the interaction of $X_1$ and $X_2$ states with the fields charged under $U(1)_X$ that receive a VEV, and induces mass splitting and mixing. In this case, $\phi$ can be a real scalar field associated with one of the Higgses responsible for the $X_1$-$X_2$ mass splitting.

\section{Conclusions}

The phenomenal success of the flavor program at high-luminosity $e^+e^-$ and hadron colliders, as well as fixed target and meson beam experiments, provides precision tests of the CKM paradigm and puts strong constraints on models of new physics. One of the most stringent tests, anticipated for many decades, is the neutrino pair production channels in the decay of the $B$ and $K$ mesons. So far, only one of such modes, the three-body decay of $K^+$ to
$\pi^+\nu\overline{\nu}$ has been observed, with limited statistics. Next generation of experiments with $K^+$, $K_L$ and $B$ mesons will detect more modes and increase precision. Because these are measurements with missing energy, other physics effects, such as production of 
dark matter particles/their mediators, may induce similar signatures~\cite{Bird:2006jd}. 

The missing energy decays of $B$ mesons have been used in the past to set limits on the pair production of dark matter states and on the single production of the Higgs-like particles. In this paper, we have addressed a possibility that $K_L$ decays can provide an additional probe. Central to our paper is the idea that $K_L$ can produce a pair of dark states, $X_{1(2)}X_2$ in the two-body decay. If one or both of these states is unstable with respect to the decay to photons or to $\pi^0$, such sequential decays will partially or completely mimic the signature of $K_L\to \pi^0\nu\overline{\nu}$. The two-body nature of the decay means that the use of $K_L$ is more sensitive to $X_{1(2)}X_2$ final states than 
$K^+$ decays, in contrast with models that modify the 
effective $d-s-\nu\nu$ vertex using the short distance physics \cite{Grossman:1997sk}. 

To be more specific, we provide several models where the $X_{1(2)}X_2$ states appear as a consequence of the $Z'$ and Higgs portals, interacting with generic dark sectors. For all models in this paper, we adopt the MFV approach, thus minimizing the number of free parameters and relating amplitudes in $B$ and $K$ decays. For both types of portals, we have found models with either scalar or fermionic $X$ states in the final states where $K_L\to X_{1(2)}X_2$ will exceed, sometimes significantly, the corresponding SM $\nu\overline{\nu}$ mode. In some models, the same type of "dark" vertex, {\em e.g.} $Z'-X_1-X_2$ governs both
$K_L\to X_{1(2)}X_2$ and $X_2\to X_1\pi^0$ decays.

Having identified the new classes of models that give enhanced $K_L$ missing energy decay signatures, we investigate the ensuing constraints and discuss the plausibility of these scenarios vis-a-vis the reported excess of events in the KOTO experiment. 
Already at the current level of sensitivity, the KOTO experiment provides strong limitations on this class of models, as demonstrated in the case studies performed in our paper. At the edge of the excluded parameter space is the region where the predicted rates can match the excess events reported recently by the collaboration. (It is evident, however, that the future progress may come only from a deeper experimental investigation into  the nature of events seen at KOTO.)

We give a detailed analysis of some benchmarks, choosing models where one and the same parameter ($G_X$, the analogue of $G_F$ for the $Z'$ mediation) governs the decays of $K_L$, $B$ mesons, and subsequent decay of $X_2$.  
By explicitly simulating the KOTO experiment, we were able to show that all pair production models can lead to large $\pt$ events that fully populate the signal region and successfully explain all observed events. This is possible even for processes with relatively small phase space, where the $X_2$ decay away from the beam is mis-reconstructed as having larger $\pt$.  We find that the $G_X \sim O(G_F)$ is currently being probed by the KOTO experiment. 
Aside of the $Z'$-mediation, we also study what appears to be a fairly minimal possibility: contact interaction of $S_1S_2$ 
scalar pair with the down-type quarks with the MFV constraint. From the point of view of low-energy phenomenology, this model has one free coupling that gives $S_1-S_2-\pi^0$ vertex that is parametrized by a small parameter $m_{eff}$. We show that KOTO provides a significant restriction on the paremeter space of the model, while $m_{eff}\simeq 100$\,eV and $m_2>400$\,MeV could in principle accommodate the excess events with no extra parameters involved. 

In all models we consider 
one should expect interesting consequences for the hadronic beam dump experiments. 
Indeed, the $c\tau$ for an unstable dark state (giving {\em e.g.} in excess of 1\,m is almost guaranteed to be ruled by the high-energy proton beam dumps. Future efforts in this direction will access $c\tau$ at the level of 10\,cm and below, 
which we find to be the most motivated range of lifetimes for $K_L\to X_1X_2\to 
\pi^0+2X_1$ scenarios. 

Finally, we address the possibility that one of the states emerging from the $K_L$ decay may be a dark matter particle. We find that $X_1-X_2-\pi^0$ vertex featured in the meson decay discussion is generally too week to produce correct dark matter abundance. However, further extension of the model based on {\em e.g.}
$X_1X_1\to {\rm unstable~particles}$ annihilation can always accommodate 
$X_1$ as dark matter (in models where it is stable). 

\vspace{1ex}
{\em Note added:} While this work was prepared for release, some related ideas were explored in Refs. \cite{Ziegler:2020ize,Gori:2020xvq}. 

\begin{acknowledgements}
The authors would like to thank Drs. A. Arvanitaki, D. Egana-Ugrinovic, K. Nakagiri, and T. Yamanaka for useful discussions and correspondence.
MP expresses his gratitude to the organizers and participants of the HC2NP 2019 workshop (Tenerife) for the intellectually stimulating environment that led to some of the ideas explored in this paper. The work by KK is supported by the  DOE grant DE-SC0011842 at the University of Minnesota. The research at the Perimeter Institute is supported in part by the Government of Canada
through NSERC and by the Province of Ontario through
MEDT.
\end{acknowledgements}

\appendix

\section{Simulation Details and Efficiencies}\label{app:pt}

We generate $K_L$ decays from the beam exit position to the surface of the ECAL, and find that $\sim 8\%$ of $K_L$ decay in the $414.8$~cm region. We do not simulate the photon conversion inside the calorimeter, but take the true incident position on the surface of the ECAL. The photon $x$ and $y$ positions and their energies are then smeared according to a Gaussian distribution with resolution as given by Eqs.~(6.27-28) of Ref.~\cite{Maeda:2016zqe}. All other reconstructed quantities are functions of these variables. Note that by not simulating the actual shower development, we cannot implement the trigger-level center-of-energy (COE) cut, as well as the shower shape cuts, but these are expected to have negligible impact on our conclusions. 

\begin{figure}[h]
    \centering
    \includegraphics[width=0.49\textwidth]{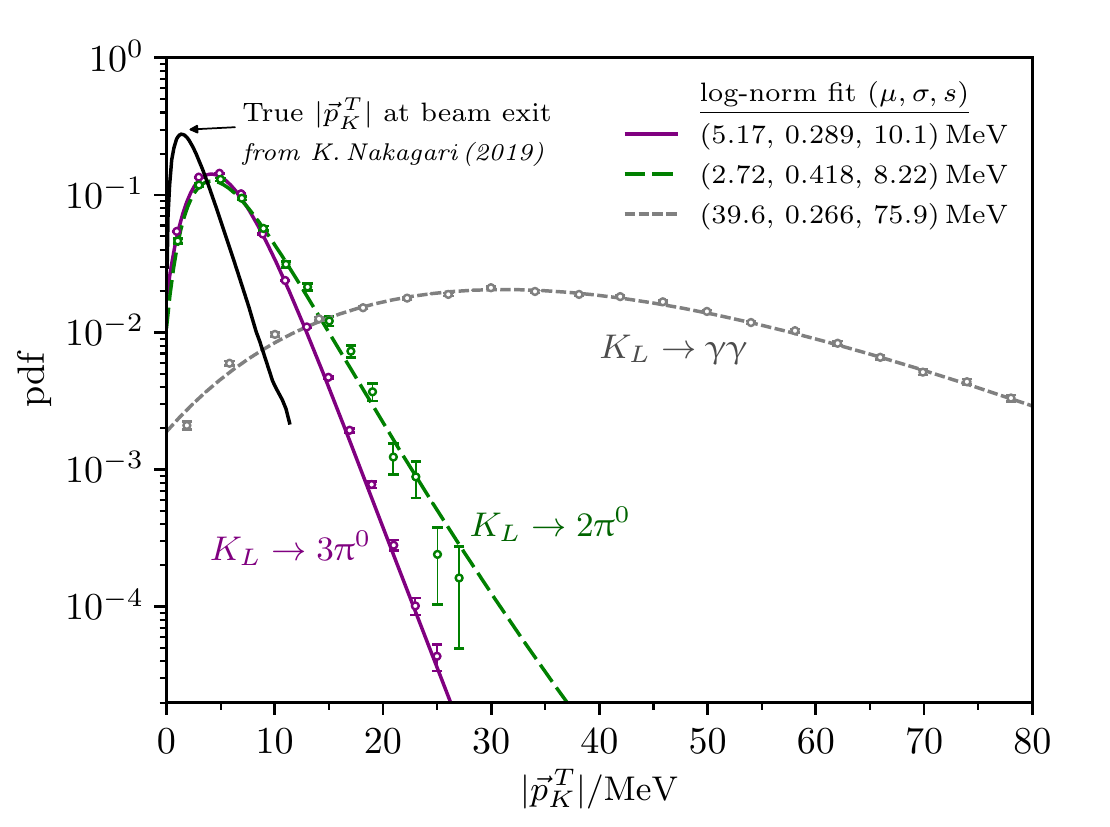}
    \includegraphics[width=0.49\textwidth]{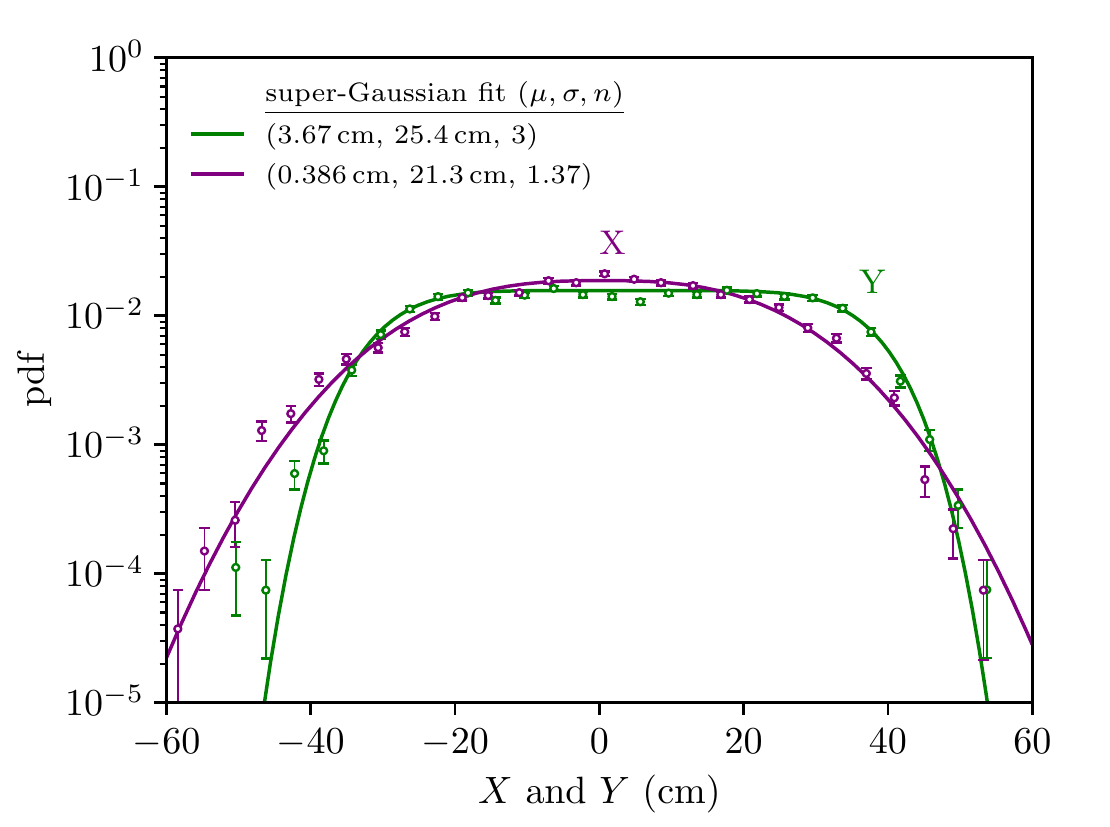}
    \caption{Area normalized distribution of $K_L$ transverse momentum at KOTO. We show the true value at the beam exit (from Ref.~\cite{Nakagiri:2019yec}) as a black solid line, and the reconstructed value as measured by the KOTO collaboration in $K_L\to 3\pi^0$ (purple), $K_L\to 2\pi^0$ (green), and  $K_L\to \gamma\gamma$ (grey) event sample. The latter measurement finds much larger values of $\vec{p}^{\, T}_{K}$ as the decay is assumed to have happened exactly along the beam line. We also show the best fit parameters of our log-normal distribution fit. \label{fig:pkpt}}
\end{figure}

Regarding the beam size, we take the $(X,Y)$ distribution as measured by $K_L\to2\pi^0$ final states. To a good approximation, the beam is a $8\times8$~cm$^2$ squared beam, but in the simulation we implement a super-Gaussian fit to the data shown in Fig.~(6.4) of Ref.\cite{Nakagiri:2019yec}. The initial transverse momentum of the neutral kaons in the beam, $\ptK$, is important as it may induce a larger $\pt$ in the signal and leads to an increase in the cross sectional area of the beam at large $Z$. To implement this in our simulation, we fit the $K_L\to2\pi^0$ data shown in Fig.~(6.3) of Ref. \cite{Nakagiri:2019yec} to a log-normal probability distribution, parametrised as
\begin{equation}
   P(\mu, \sigma, s) \equiv  \frac{1}{(x+\mu) \,\sigma\, \sqrt{2\pi}} e^{- \frac{\ln^2{\hat{x}}}{2\sigma^2}}, \quad \text{where}\quad \hat{x} = \frac{x+\mu}{s}.
\end{equation}
The prediction and the fit are shown in \reffig{fig:pkpt}. For comparison, we also plot the other KOTO normalisation data as a function of $\pt$ from $K_L\to 3\pi^0$ and $K_L\to\gamma\gamma$ samples. The severe broadening of $\ptK$ observed in $K_L\to\gamma\gamma$ comes from the assumption that the decay has occurred exactly along the center of the beam $(X,Y)=(0,0)$, similarly to the procedure for $K_L\to\pi^0\overline{\nu}\nu$ signal. For the multi-pion final states, however, there are enough constraints to reconstruct the full kinematics, and the previous assumption is not necessary.

As a sanity check of our KOTO simulation, we show in \reffig{fig:comparisons} the distributions obtained with SM $K_L\to\pi^0\overline{\nu}\nu$ decays, showing reasonable agreement with the KOTO Monte Carlo curve as taken from Fig.~(6.17.c) of Ref.~\cite{Nakagiri:2019yec}.

\begin{figure}[h]
    \centering
    \includegraphics[width=0.49\textwidth]{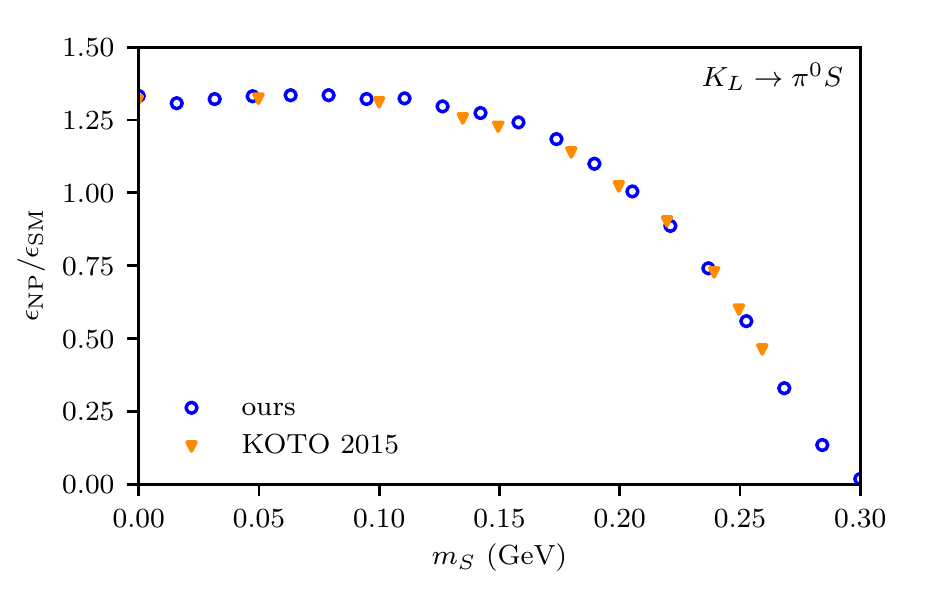}
    \includegraphics[width=0.49\textwidth]{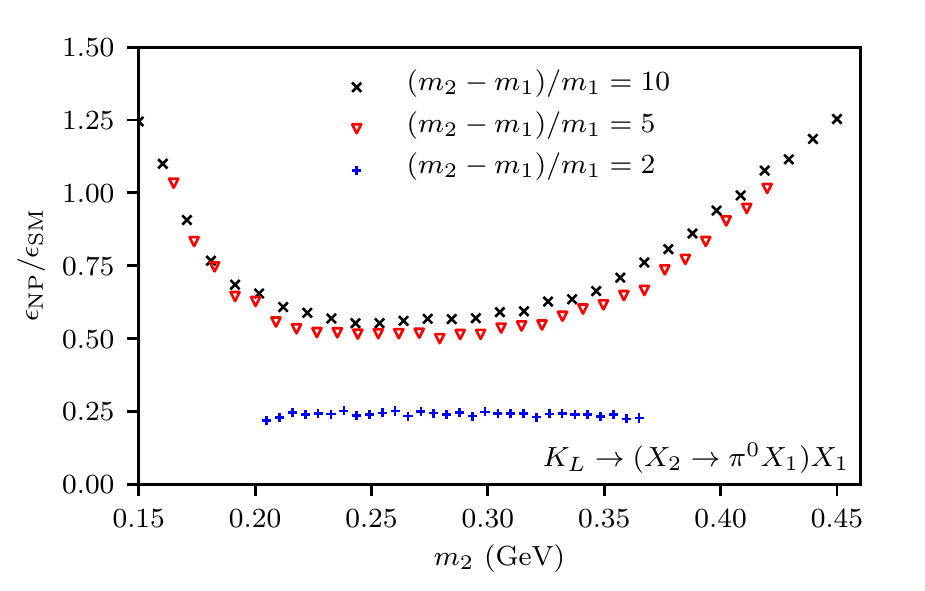}
    \includegraphics[width=0.49\textwidth]{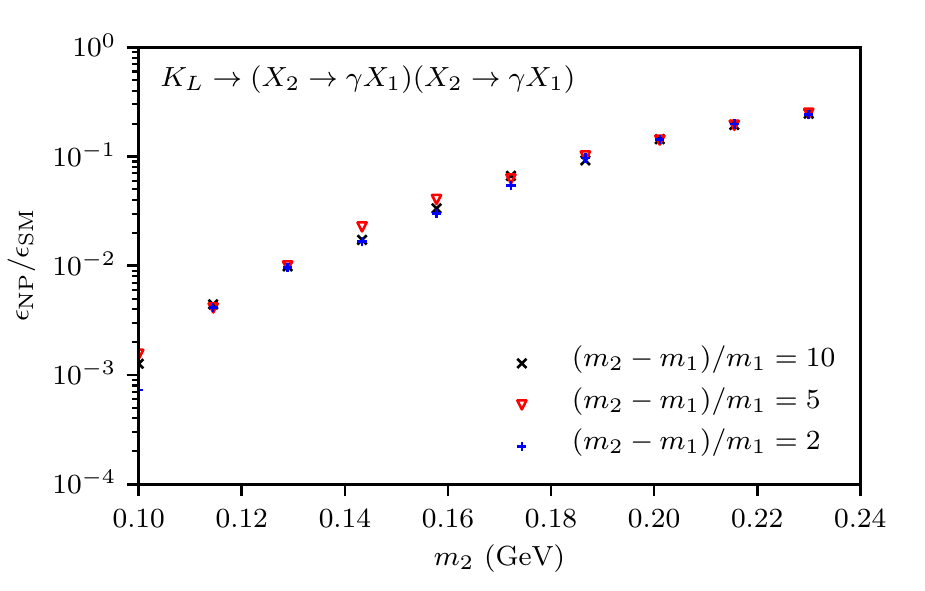}
    \includegraphics[width=0.49\textwidth]{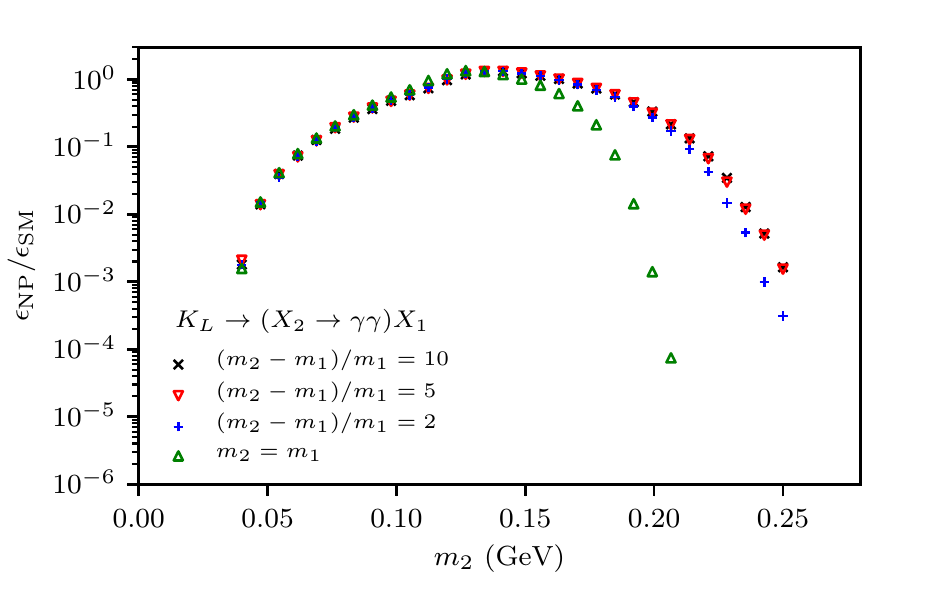}
    \caption{The ratio between signal selection efficiencies in our new physics ($\epsilon_(\rm NP)$) and in the SM ($\epsilon_{\rm SM} = \epsilon (K_L\to \pi^0 \overline{\nu} \nu)$). We take all new physics decays as prompt. \textbf{Top left:} the singlet scalar $S$ case for validation. \textbf{Top right:} the $\pi^0$ prodcution scenario (A). \textbf{Bottom left:} the dipole portal scenario (B).  \textbf{Bottom right:} the $\pi^0$ impostor scenario (C).  \label{fig:efficiencies}}
\end{figure}
We also compute the ratio of signal selection efficiencies between our new physics signals and the SM $K_L\to \pi^0 \overline{\nu}\nu$ rate, all within our simulation. These are shown in \reffig{fig:efficiencies}. We show the case of $K_L\to \pi^0 S$ to validate our simulation, where we implemented the cuts on ($\pt$,$Z_{\rm vtx}$) as used in the 2015 analysis. In quoting the 2015 KOTO values, we assume that the ratio between the upper limit on invisible $K_L\to \pi^0 S$ and $K_L\to \pi^0 \overline{\nu}\nu$ processes, taken from Fig.~4 of Ref.~\cite{Ahn:2018mvc}, is equal to the ratio of signal selection efficiencies. All other plots include the new 2016-2018 cuts on ($\pt$,$Z_{\rm vtx}$).

\section{Three-body decays}\label{app:threebody}

Focusing on a $Z-Z^\prime$ mixing scenario, we now show the analytical expressions for pair production of dark states in three-body decays of $K^+$ and $B$ mesons, as well as the invisible three-body decays for $X_2$. These $M^\prime\to M$ transitions proceed via vector current in the $Z^\prime$ portal
\begin{equation}
    \langle M(p)|\overline{Q} \gamma_\mu Q^\prime | M^\prime(p^\prime)\rangle = (p^\prime_\mu+p_\mu) f_+(q^2) +\frac{m_{M^\prime}^2-m_M^2}{q^2} (p^\prime_\mu-p_\mu) 
    [f_0(q^2)-f_+(q^2)],
\end{equation}
and via scalar transitions in the scalar portal
\begin{equation}
    \langle M(p)|\overline{Q}  Q^\prime | M^\prime(p^\prime)\rangle = \frac{m_{M^\prime}^2 - m_M^2}{m_{Q^\prime} - m_{Q}} f_0(q^2)
\end{equation}
where $q=p^\prime-p$ and $Q,Q^\prime \in \{b,s,d\}$ depending on what transition is considered. Neglecting the final state masses and treating the form factor as a constant ($f_+\simeq0.33$ for $B\to K$, and $f_+\simeq 0.97$ for $K^+\to \pi^+$ transitions), we find
\begin{equation}
\label{simple1}
   \Gamma_{M^\prime\to M \psi_1 \psi_2}^{\rm D} \approx  \frac{|g_{Q^\prime Q X}|^2 g_X^2(|c_V|^2+|c_A|^2)}{1536\pi^3} \frac{m_{M^\prime}^5}{m_{Z^\prime}^4} f_+^2,
\end{equation}
for the Dirac fermions case. For Majorana fermions, the rate above is multiplied by an additional factor of two. For a $Z^\prime$ coupled to a scalar-pseudoscalar pair, we find  
\begin{equation}
\label{simple2}
   \Gamma_{M^\prime\to M S_1 S_2} \approx  \frac{|g_{Q^\prime QX}|^2 {g_X^2}}{6144\pi^3} \frac{f_+^2 m_{M^\prime}^5 }{m_{Z^\prime}^4}.
\end{equation}
 In the full $B$ meson decay computation, we make use of the form factors in Ref.~\cite{Ali:1999mm}, and in $K^+$ meson decay we use the Taylor-expanded dispersive parameterization in Ref.~\cite{Carrasco:2016kpy}.

\begin{figure}
    \centering
    \includegraphics[width=0.49\textwidth]{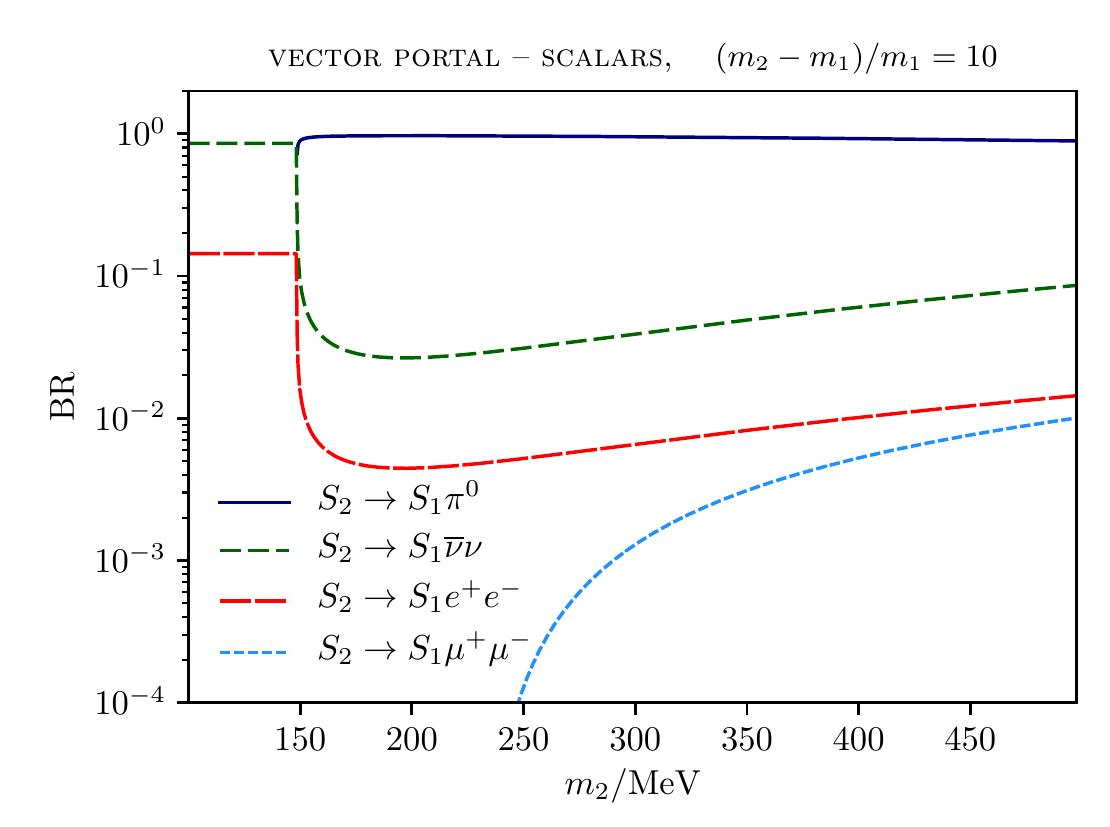}
    \includegraphics[width=0.49\textwidth]{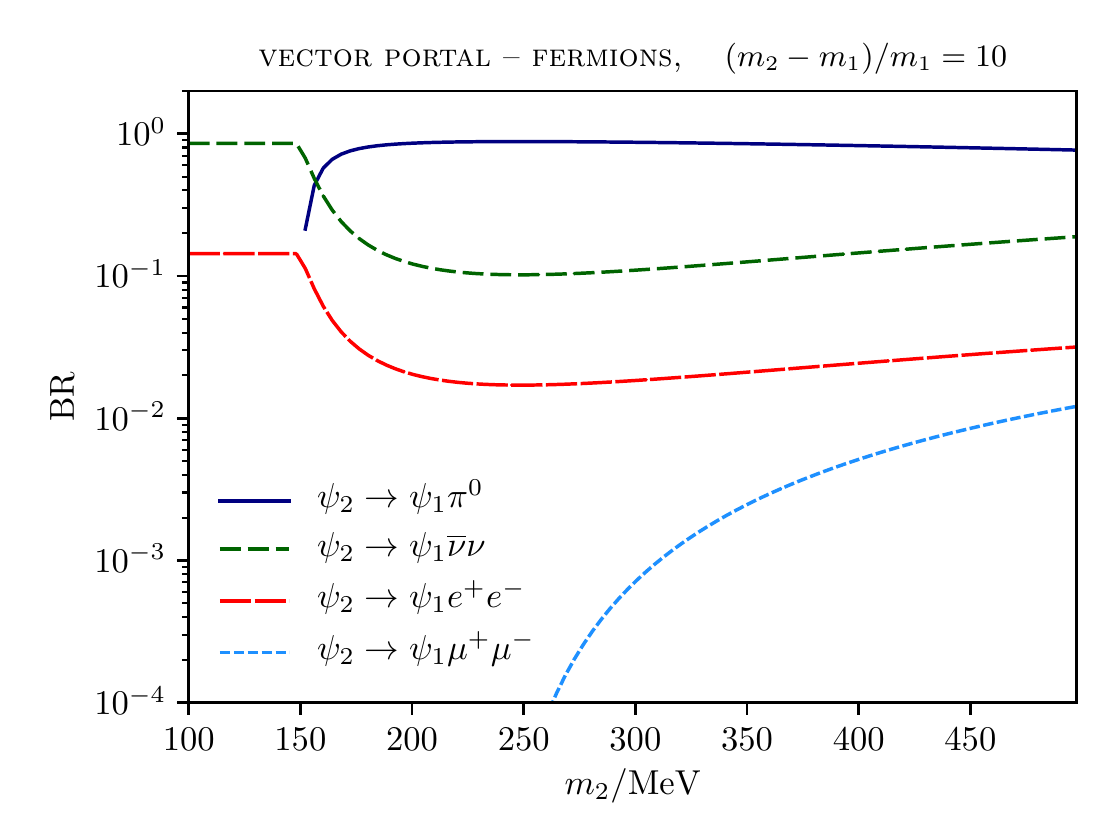}
    \caption{Branching ratios for $X_2$ particles in the vector portal model with with scalars (left) and fermions (right).\label{fig:X2BR}}
\end{figure}

Now, we compute the heavier dark state three-body decays into SM fermions. This is relevant as the decays to $X_2\to X_1\overline{\nu}\nu$ set an intrinsic invisible decay $\rm{BR}$ for $X_2$. For the Dirac fermion case, neglecting the SM fermion mass, we find
\begin{align}
\Gamma_{\psi_2\to\psi_1 \overline{f}f}^{\rm D} = \frac{G_X^2 m_2^5}{192 \pi^3} \left( |c_V|^2 G(-y_1)+|c_A|^2 G(y_1)\right) \left( (g_V^f)^2+(g_A^f)^2\right)
\end{align}
where $y_a=m_a/m_2$, $G(x)=1+2x-8x^2+18x^3-18x^5+8x^6-2x^7-x^8 +24 x^3(1-x+x^2)\log{|x|}$, and $g_{V,A}^f$ are the SM vector and axial weak couplings of $f$. For Majorana fermions, the previous decay rates are larger by a factor of two. It is clear that the decay into $\pi^0$ typically dominates, with a BR of order $80\%$ provided $m_1\ll m_2-m_\pi$. For the scalar decay case, we find
\begin{align}
&\Gamma \left( S_2 \to S_1 \overline{f} f\right)=\frac{G_X^2 m_2^5}{384 \pi^3} (g_V^{f\,2} + g_A^{f\,2})  H(y_1),
\end{align}
where $y_a=m_a/m_R$, and $H(x)= 1-8x^2-24 x^4 \log{x}+8x^6-x^8$. A full computation including the fermion mass gives the BR in \reffig{fig:X2BR}, and shows the intrinsic $\lesssim 10\%$ invisible BR.

\bibliographystyle{apsrev4-1}
\bibliography{main}{}

\end{document}